\documentclass[11pt,a4]{article}
\usepackage{amsmath}
\usepackage{amssymb}
\usepackage{graphicx}
\usepackage{subfigure}
\usepackage{listings}
\usepackage[usenames]{color}

\usepackage[pdftex,
      colorlinks=true,
      urlcolor=blue,      % \href{} local file
      filecolor=blue,      % \href}{} external (URL)
      linkcolor=blue,       % \ref} and \pageref{}
      citecolor=blue,
      pdfauthor={Georgios Choudalakis},
      pdfpagemode=None,
      bookmarksopen=true]{hyperref}

\definecolor{MyBkgColor}{rgb}{0.99,0.99,0.90}
\definecolor{commentcolor}{rgb}{0.5,0.3,0.93}

\begin{document}

\title{How to Use Experimental Data to Compute the Probability of Your Theory}
\author{Georgios Choudalakis \\ {\it \small University of Chicago, Enrico Fermi Institute,} \\ {\it \small 5640 South Ellis Avenue. Chicago, IL 60637} \\ \it{ \small CERN office: B40-R C32.}}
\date{\today}

\maketitle

\begin{abstract}
This article is geared towards theorists interested in estimating parameters of their theoretical models, and computing their own limits using available experimental data and elementary Mathematica$^\copyright$ code.  The examples given can be useful also to experimentalists who wish to learn how to use Bayesian methods.  A thorough introduction precedes the practical part, to make clear the advantages and shortcomings of the method, and to prevent its abuse.  The goal of this article is to help bridge the gap between theory and experiment.
\end{abstract}

%\tableofcontents

\section{Introduction}

The intention of this article is to make it clear to theorists how to use available experimental data to carry out standard Bayesian inference, by which they can estimate and set limits on parameters of interest of their own theoretical models or, depending on the mood, of their friends' models.

\paragraph*{Disclaimer:} The method described in this article is not, currently, an official recommendation of any experimental collaboration.  The author is a High Energy experimentalist who writes on his own behalf.  The strengths and limitations of the method will be explained, so, the readers should use their own judgement, as always.

\subsection{Warning}

One should never think it's possible to claim a discovery without consulting with the experimental collaboration which produced the data.  If one suspects something significant is seen in some data, it is essential to investigate possible detector effects, or other experimental factors that could explain it, before attributing it to new physics.   

Interpretations should be discussed with the experimentalists who produced the data you use!  {\em The goal of this article is to strengthen the collaboration between theorists and experimentalists, not to let theorists run off with potentially wrong conclusions.}

\subsection{Why Bayesian inference?}

Bayesian inference dates back to the 18th century, and is based on solid theoretical ground: standard probability theory, which underpins Bayes' theorem.  So, although in the last years it has emerged as the cutting edge of Statistics, it is actually very old.

Luckily, Bayesian inference is very easy to carry out, which makes it possible to propose here a practical procedure that theorists can actually use without complicated software or large computing power.

The fundamental advantage of Bayesian inference, compared to Frequentist methods,  is that it makes statements directly about the {\bf parameter of interest (POI)}.  Namely, in the end one finds the {\bf probability density function (PDF)} of his POI.  This is not true for any of the Frequentist constructions, where confidence intervals are obtained.  There, no statement is made about the probability of the POI to be within any interval. The coverage of a Frequentist confidence interval is a statistical property of the confidence interval itself, {\em not} the probability for the POI to be within that interval, as many wrongly think.  The Frequentist approach is to assume various values for the POI, and compute how likely the data would be according to each value, and then set the limit at the value which, if assumed, starts making the observed data very unlikely.  But the fact $P(\rm data | hypothesis)$ is small doesn't entail that also $P(\rm hypothesis | data)$ is small, and what one really asks is the latter\footnote{Think of the classic example:  The obvious difference between $P(\rm pregnant|female)$ and $P(\rm female|pregnant)$.}.  The Bayesian approach is to assume the data, and find how likely each hypothesis is, thus compute $P(\rm hypothesis | data)$ directly.

\subsection{The prior}
\label{sec:prior}

A necessary ingredient of inference is the prior, which is the PDF assumed for the POI before seeing the data.  This prior PDF could be the posterior of a previous experiment, but the latter too would depend on some prior used to interpret the data of the previous experiment.  In the end it is impossible to avoid the dependence on some prior\footnote{This very first prior, which is used to make an inference with the very first data, is sometimes called "uberprior".  As more data is accumulated, the uberprior keeps being updated to give newer posteriors.  It makes no difference if the uberprior is updated with all experimental observations at once, or if the updates are done incrementally using the posterior of the last inference as a prior for the next inference.}.  

Some people, the so-called ``subjective Bayesians'', embrace the prior as a means to express the mathematical fact that the conclusion (i.e.\ the posterior PDF) doesn't depend only on the evidence (i.e.\ the data), but also on the initial assumptions under which this evidence is interpreted (i.e.\ the prior PDF).  Not surprisingly, different people will arrive to different conclusions, or, the same person will arrive to different conclusions, if he starts from different assumptions.  These assumptions don't have to reflect anyone's subjective distribution of probability, since nobody prohibits asking what the result would be if the prior was different, regardless of personal preferences.

 To draw an analogy, the posterior is like a function ($f$) of the prior ($x$).  Just like the function $f(x)$ could be evaluated at any $x$, a posterior can be computed for any prior.  In the case of a function mapping $\mathbb{R} \to \mathbb{R}$, it is easy to plot $f(x)$ versus $x$ to visualize the function.  Unfortunately, this can not be done on a piece of paper when $x$ is a prior PDF and $f(x)$ is a posterior PDF.   Still, it should be possible to plug in a prior and easily evaluate the corresponding posterior.  This convenience is offered by the method presented here.

It is not surprising that the posterior depends on the prior, and it doesn't mean that the results are not well-defined.  For each prior, the posterior is unique, and determined by the data.
Ultimately, with more data, all priors asymptotically result in the same posterior, except for very extreme priors, like the Kronecker $\delta$ function which represents an unshakeable prior conviction.  
The prior allows to interpret the same data from various starting points, including even the interpretation of someone with an unshakeable prior conviction.  In this sense, any prior is legitimate; even a Kronecker $\delta$, although most wouldn't find interesting the inferences of someone who was committed to note letting data change his mind.  That's why every Bayesian result must be accompanied by a statement of the assumed prior, to know how to judge it.

Other people, the so-called ``objective Bayesians'', view the prior as something undesirable.  Since it is impossible to eliminate, they try at least to prescribe its definition.   There are prescriptions which offer the posterior specific properties that some consider important, such as independence of the result under re-parametrization of the POI.  Other prescriptions try to achieve the opposite effect of a Kronecker $\delta$, namely, maximal susceptibility to the data. % which is quantified by maximizing the expected change in entropy between the prior and the posterior.
 To use the previous analogy, these efforts are like prescribing a value of $x$ with some special property; for example the $x$ which maximizes $\frac{df(x)}{dx}$.  Some would argue that, if we can't plot $f(x)$ for all values of $x$, let's at least compute $f(x)$ at that special $x$ that has some (subjectively?) interesting property.  A couple of criteria for this were mentioned already.
 
For a ``subjective Bayesian'', since all priors are fine, so are these special priors, which are known as ``non-informative'' priors.  It should be mentioned, though, that if one follows the prescription to compute a non-informative prior (which can be quite cumbersome), he may not be satisfied with the result, because it is highly unlikely to reflect any intuitive guess anyone would have made for the POI.  Such priors lose their meaning as distributions of prior belief, and become {\em devices} used to {\em tune} the properties of the posterior.  For example, non-informative priors often depend on the expected background.  To see how paradoxical this is, consider that, if an experimental device registered more background noise for any instrumental reason, we would have to change accordingly our prior PDF of the Higgs mass, or some other fundamental POI that the device would be supposed to measure.  One would think that our prior PDF for the Higgs mass should have nothing to do with how much background is registered by some instrument.  But again, if that is the prior an ``objective Bayesian'' wants to try, a ``subjective Bayesian'' has no reason to object.  The method presented here allows the readers to plug in any prior they wish, including even non-informative priors.%, if they know how to compute them, or if they find them in the literature.

\subsection{Systematic uncertainties}

This article shows how to set an {\em exact} limit to your own signal, ignoring systematic uncertainties.
The basic principles of including systematic uncertainties, with some examples, will be given in Section~\ref{sec:systematics}.  It is not possible, however, to provide a complete general prescription for this, because not all systematic uncertainties are the same.  The reader will have to generalize a little the examples provided here.  Theorists are equipped to evaluate theoretical systematic uncertainties, but experimental uncertainties are the expertise of experimenters.  Collaboration is necessary for a complete result.

Most theorists would be satisfied with limits which ignore systematic uncertainty, since they are typically only a few per-cent different from the limits with full treatment of systematic uncertainty.  Given that it is practically impossible for the experimentalist to compute limits to all possible theories of the present and the future, it is important for theorists to be able to easily set limits, even with the approximation of ignoring some systematic uncertainties.  An approximation is better than nothing.  

Furthermore, if an experiment uses a benchmark model to demonstrate the impact of systematic uncertainties, that can be used as a guideline to estimate the impact of the same uncertainties on another model,  though for some uncertainties the impact may depend on the signal.

If someone has a reliable model of systematic uncertainties, Section~\ref{sec:systematics}, will allow him, in principle, to convolute these uncertainties.  %However, one needs to know that this is not always simple.

\subsection{Modeling the detector response}
\label{sec:detectorSimulation}

This article is not trying to address the issue of detector simulation.  It is assumed that the theorist can approximate the distribution of his signal after reconstruction.  Many theorists do this with tools like PGS \cite{PGS}.  Experiments also provide their acceptance to objects (jets, leptons) as a function of quantities accessible to theorists, such as transverse momentum ($p_T$) and pseudo-rapidity ($\eta$).  In some cases\footnote{For example, in \cite{ATLASdijetPLB}, the amount of detector smearing in dijet mass is given, so, a hadron-level dijet mass value can be smeared, stochastically, to model the detector-level dijet mass.} the detector resolution is also parametrized, so, a theorist can approximately smear the energy of jets and leptons.

It is often claimed that unfolding \cite{glenUnfolding} the experimental data allows theorists to test their theories without needing detector simulation.  This is an idealization of the actual situation \cite{myUnfolding}.  There are many ways to do unfolding; it is not as unique and well-defined as the data.  Regularization, which plays central role in unfolding, depends on some arbitrary choices.  The root of all problems with unfolding is that it is impossible to recover information that is lost during detector smearing.  The unbiased estimator of the true spectrum has enormous variance, which makes it useless, so during regularization one introduces some bias, on purpose, to reduce the variance.  In practice, unfolding may introduce more difficulties than it solves, so, it's advisable to avoid it unless nothing else is possible.  Unfolded spectra are estimators that follow complicated probability distributions; it is no longer correct to treat each bin as an independent observation, or to assume that its contents follow a Poisson or Gaussian distribution.  So, simple tests like $\chi^2$/(degrees of freedom) are no longer correct.  There are bin-to-bin correlations which are usually not published.  Even if a correlation matrix is provided, it assumes that the multidimensional PDF of the estimator is Gaussian, but in reality its shape is irregular, especially when low statistics appear in some bins.   The bias that is introduced by regularization is typically larger in parts of the spectrum where statistics are lower, which is precisely where exotic effects might be.  In reality it is impossible to estimate the actual bias of unfolding, unless we knew the actual spectrum of the data before smearing, which is obviously unknown, and if we look for new physics it can not be assumed to be given by Standard Model (SM) simulation prior to smearing, or else we wouldn't be looking for new physics.  So, searches for new physics are an unfavorable environment for unfolding.  

If an experimentalist has a matrix of migrations, which is the main ingredient of all unfolding methods, it is better to publish the matrix to allow theorists to fold the detector smearing into their theoretical signal, instead of using the matrix to unfold the data.  This works without problems because, while it is impossible to recover lost information, it is totally possible to reduce existing information.  The benefits of smearing, or folding, compared to unfolding, are the following:  (a) Unlike data, the theoretical prediction before smearing does not have statistical fluctuations, so there is no need for regularization.  A simple multiplication of the folding matrix with the spectrum prior to smearing returns the expected spectrum after detector smearing.  (b) Since the data are not unfolded, they follow a well-known PDF, e.g.\ Poisson, Binomial, or Gaussian.  Each bin can be used as an independent observation, so, there is no need to consider complicated (and inevitably approximate) correlations among data in different bins.  It is simple to compare the data to the folded theoretical spectrum using simple methods such as a $\chi^2$ or a likelihood test.

While folding solves some of the problems of unfolding, it faces a difficulty:  Different theoretical models would require different folding matrices to be folded correctly\footnote{The same problem exists to some degree in unfolding.  The elements of the migration matrix, which are usually derived by passing the standard model prediction through detector simulation, are not realistic if there is new physics.}.  To see why, imagine new particles of different spin, whose decay products would be distributed differently in $\eta$, thus measured by different parts of the detector, thus suffering different amounts of smearing.  That is why it is difficult to provide folding matrices that would work equally well with all theories.  Probably the best strategy is to model detector effects in the level of measurable objects (jets and leptons).  By smearing each object separately, we can smear any signal that decays into such objects.

This article allows a theorist to easily assume different signal distributions, therefore, if there is some doubt about the exact signal shape after detector smearing, it is easy to try different possibilities.  The data, however, have to always be the observed data, {\em not} the output of any unfolding.  If an experimental analysis chooses to use unfolding, always ask also for the original data, because {\em there}  one can see reality; unfolding offers mere interpretations.

\subsection{Analysis event selection}
\label{sec:selection}

To model the signal that makes it to the final plot, it is necessary to apply the event selection of the analysis whose data are used.

Analyses always publish the event selection they apply.  Typically, the selection applies to single objects, or combinations of objects.   For example, ``cuts'' are made in transverse momentum ($p_T$), pseudorapidity ($eta$) or rapidity ($y$), differences in azimuthal angles $(\Delta \phi)$, differences in rapidity or pseudorapidity, scalar or vectorial sums of transverse momenta, and missing transverse energy (MET).  

A theorist has easy access to the 4-vectors of quarks, gluons, leptons, and to exotic particles escaping detection (e.g.\ stable or long-lived neutralinos).  With generators like {\sc Pythia} \cite{Pythia}, it is also possible to have access to hadronic showers resulting from emitted quarks and gluons, and using jet clustering algorithms such as those implemented in {\tt FastJet} \cite{fastjet} it is possible to define hadron-level jets.

For leptons it is simple to apply kinematic cuts.  For jets it is a little less simple.  If a theorist has hadron-level jets, their energy corresponds (on average) to the energy of calibrated reconstructed jets on which experimentalists typically apply $p_T$ cuts.  So, it is acceptable to apply the same $p_T$ cut to hadron-level jets.  The situation is a little more tricky when one has access only to partons, before showering and hadronization.  In that case, one needs to consider that the hadron-level $p_T$ differs from the parton $p_T$, due to out-of-cone losses.  To apply a lower $p_T$ at hadron level, a slightly higher threshold is necessary at parton level.  This becomes less of an issue when a large jet size parameter is used, or if the partons (and jets) are at higher $p_T$, thus more boosted, thus having less out-of-cone losses.  For anti-$k_T$ jets with $R$=0.4, the out-of-cone energy fraction is roughly 10\% at hadron-level $p_T \simeq 30$~GeV, it reduces to about 4\% at 100 GeV, and is less than 2\% at $p_T>200$~GeV.  For anti-$k_T$ jets reconstructed with R=0.6, the out-of-cone fraction is less than 2\% even at $p_T \simeq 30$~GeV.  Usually jets produced by new physics carry large momentum, well above such $p_T$ thresholds, so such differences wouldn't be important.

When a minimum MET is required, this translates into a minimum $p_T$ carried by the vectorial sum of neutrinos, gravitons, neutralinos etc.\ in the signal.  Similar to charged leptons, one has the momenta of these objects, so it is simple to add vectorially their transverse momenta and apply the same threshold.  In reality, there is some MET even if at parton level all objects balance perfectly.  That (fake) MET comes mostly from the finite energy resolution of the calorimeter, from detector cracks, beam remnants, etc.  Detector simulation reproduces this MET.  It can be approximated by smearing, according to detector resolution, the transverse momenta of any jets or other objects in each event.  However, fake MET is typically much less than the real MET produced by exotic particles that escape detection, and it is also well below the MET thresholds required in searches for such particles.  So, it should be safe to ignore fake MET when genuine MET is part of a new physics signature.

\subsection{Where to find data}
\label{sec:dataSources}

A source of data is the HepData project, hosted at the University of Durham \cite{HepData}.  From there, anyone can retrieve the observed and expected spectrum in bins of specified delimiters, for an increasingly number of analyses from various experiments.

Other disciplines, such as observational astrophysics, have a culture of open access to data.  Since the author is a High Energy experimentalist, the focus of this article is in High Energy physics, but it must be obvious how the methods shown here can be used in other disciplines.

As many theorists evidently know, there is software which enhance one's ability to read numbers off of published figures.
An example is {\tt DataThief} \cite{DataThief}.  Don't let the name of this software fill you with guilt;  there is nothing unethical in reading values off of a published and peer-reviewed experimental plot.  Just beware that, due to limited resolution, it may be hard to ``see'' exactly the content and the delimiters of each bin.  It can't hurt to ask an experimental collaboration to publish the exact values, to avoid approximations.

\subsection{Software used}

Since most High Energy theorists are familiar with Mathematica\footnote{Mathematica$^\copyright$ is proprietary software, developed by the Wolfram Research company.}, we will use it here for demonstration.  Very elementary use of Mathematica is made, so, anyone should be able to read the code shown here and understand what it does.

The computation is so simple that, with some perseverance, it can be carried out even by hand.  No generation of Monte Carlo events is required for Bayesian inference.  Mathematica provides an intuitive, interpreted language, that makes easy also the visualization of results.

Of course, once the computation method is understood, it can be ported to any programming environment.

\subsection{The numbers used}

For the purposes of this article, it doesn't matter where the data come from.  They can be considered fictitious.  

\section{Poisson-distributed data}
\label{sec:poisson}

Let's start with the very common case of a spectrum where events are counted in defined intervals (bins) of some observable.  As a representative example, consider the distribution of events in some measured mass, as was done in \cite{ATLASdijetPLB, CMSdijet} and numerous other analyses.

Let $d_i \in \mathbb{N}$ denote the number of events observed in bin $i$, and $b_i \in \mathbb{R}^\ast$ the number of events expected in bin $i$ if there is no new physics.  The index $i$ runs from 1 to $N$, which is the total number of bins.  Let $s\in \mathbb{R}$ be the expected number of produced signal events, {\em which is our POI}.  One would simply divide $s$ by the integrated luminosity to transform it to the cross-section of the new physics process.   Let $f_i \in \mathbb{R}$ be the fraction of the produced signal that ends up in bin $i$ after detector reconstruction and event selection.  By definition, the total signal acceptance (times reconstruction efficiency) is $A$:
\begin{equation}
  \sum_{i=1}^N f_i = A.
  \label{eq:acceptance}
\end{equation}
If $A = 1$, then all signal makes it into the $N$ bins of the spectrum after event selection.  If $A<1$, then some of the signal doesn't make it to the final spectrum, either due to detector inefficiency, or because it fails some of the analysis cuts.  In either case, the array of $f_i$ values completely determines the expected signal distribution after detector smearing and event selection.  One uses $f_i$ to specify his model.  To do so one needs to calculate the theoretically predicted signal distribution in $N$ bins, and model the effect of detector smearing (Section~\ref{sec:detectorSimulation}) and event selection (Section~\ref{sec:selection}).

If $s$ signal events are produced, then the expected events in bin $i$ are $b_i + s\cdot f_i$.  Assuming that the data are indeed the data, and not the product of some unfolding (see Section \ref{sec:detectorSimulation}), the likelihood of the data under the assumption of $s$ produced events, which are distributed according to $f_i$, is:
\begin{equation}
L({\rm data}|s) = \prod_{i=1}^N {\rm Poisson}(d_i | b_i + s\cdot f_i) = \prod_{i=1}^N \frac{(b_i + s\cdot f_i)^{d_i}}{d_i!} e^{-(b_i+s\cdot f_i)}.
\label{eq:likelihood}
\end{equation}

Applying Bayes' theorem, the posterior PDF\ of our POI ($s$) is
\begin{equation}
p(s|{\rm data}) = L({\rm data}|s)\frac{\pi(s)}{\mathcal N}, 
\label{eq:posteriorPoisson}
\end{equation}
where $\pi(s)$ is the prior PDF\ (see Section \ref{sec:prior}), and ${\mathcal N}$ is a constant which normalizes the posterior to 1:
\begin{equation}
{\mathcal N} = p({\rm data}|s) = \int L({\rm data}|s) \pi(s) ds.
\label{eq:N}
\end{equation}

\begin{figure}
\subfigure[]{\includegraphics[width=0.5\textwidth]{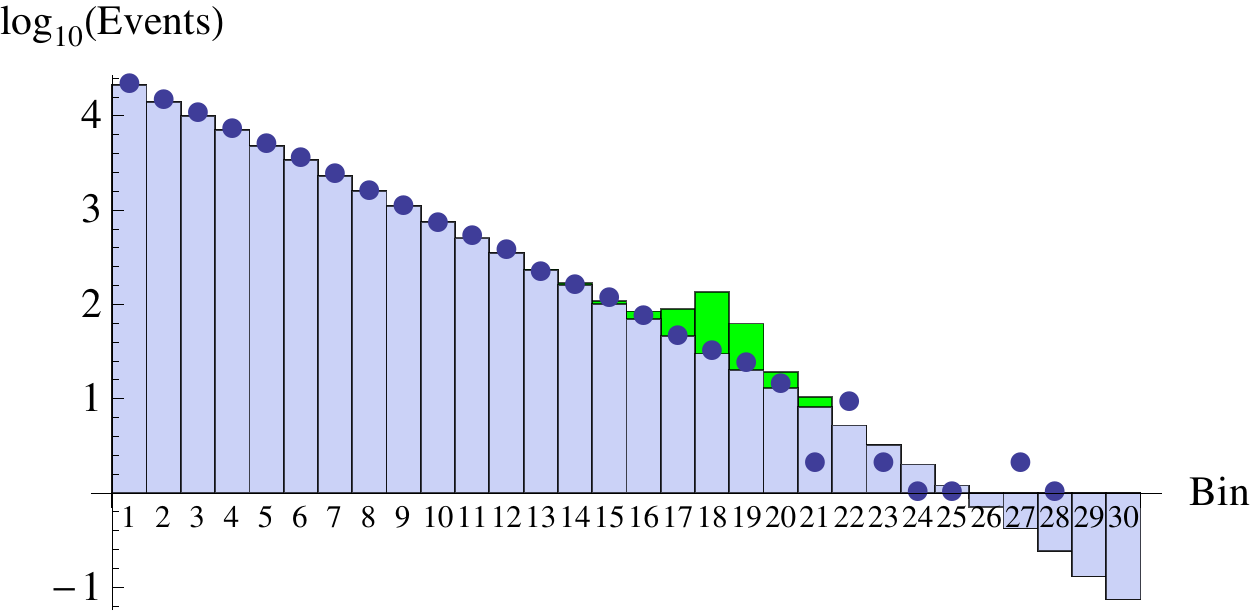}}
\subfigure[]{\includegraphics[width=0.5\textwidth]{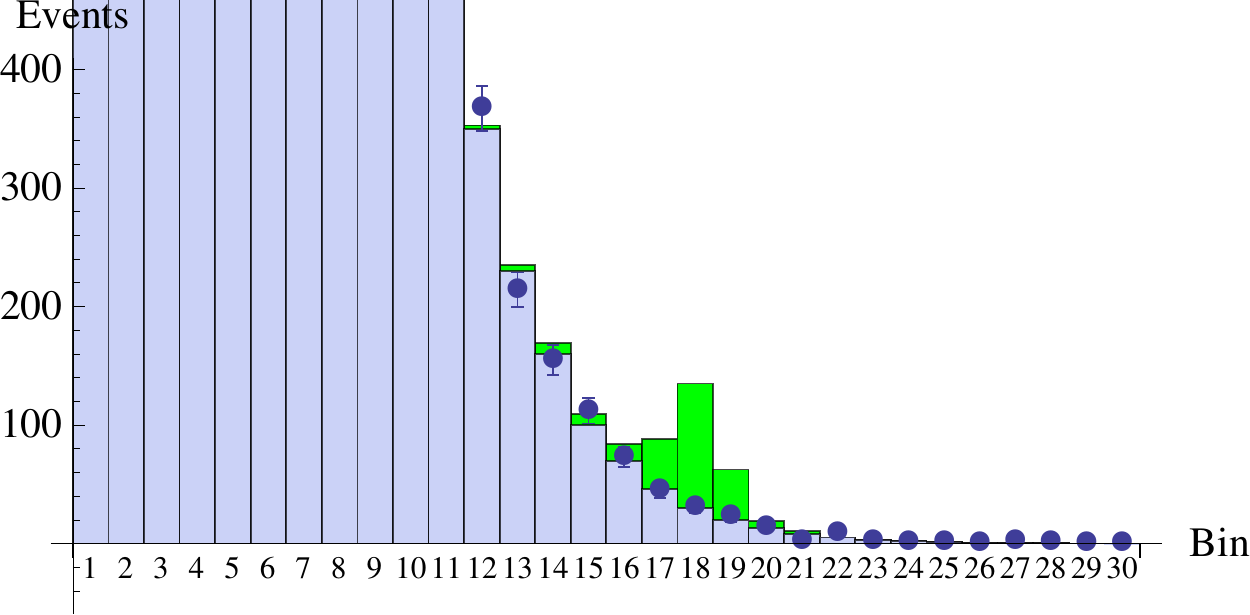}}
\caption{An example of data (markers), background (blue bars), and assumed signal distribution for $s=500$ (green bars stacked on top of background), in linear (left) and logarithmic scale (right).  The numbers have been chosen to resemble a mass spectrum like those studied in resonance searches.  \label{fig:data1}}
\end{figure}

Let's use some numbers. The data are represented by the array $d$, and the background by the array $b$, and the signal distribution by $f$, each with $N=30$ elements:
\begin{lstlisting}
d = {20839, 14404, 10285, 7094, 4841, 3440, 2338, 1555, 1059, 706, 515, 367, 214, 155, 112, 73, 45, 31, 23, 14, 2, 9, 2, 1, 1, 0, 2, 1, 0, 0}; 
b = {21000., 14000., 10000, 7100., 4800., 3400., 2300., 1600., 1100., 740., 500, 350., 230., 160., 100, 70., 46., 30., 20., 13., 8.2, 5.2, 3.2, 2.0, 1.2, 0.71, 0.42, 0.24, 0.13, 0.074};
f = {0, 0.0000105, 0.000335, 0.000485, 0.00015, 0.0008, 0.00115, 0.00425, 0.0022, 0.0034, 0.00495, 0.0055, 0.0095, 0.018, 0.0185, 0.028, 0.085, 0.21, 0.085, 0.0125, 0.0044, 0, 0.0000105, 0, 0, 0.000335, 0, 0, 0, 0};
\end{lstlisting}
The above inputs are plotted in Fig.~\ref{fig:data1}.  

The following lines compute the posterior of eq.~\ref{eq:posteriorPoisson}:
\begin{lstlisting}
nBins = Length[b] (*d,b & f should all have the same length*)
A = Total[f]      (*The acceptance*)
L[s_] := Exp[Sum[d[[i]]*Log[b[[i]]+s*f[[i]]],{i,1,nBins}]-s*A]
Prior[s_] := UnitStep[s]
NormConst = NIntegrate[L[s]*Prior[s],{s,-Infinity,Infinity}]
Posterior[s_] := L[s]*Prior[s]/NormConst             
\end{lstlisting}

\begin{description}
\item[Line 1:] simply identifies the number of bins from the size of the $b$ array.  The variable {\tt nBins} is 30 in this example.
\item[Line 2:] The acceptance is computed, as the sum of the elements of the $f$ array.  In this example, $A\simeq 0.49$.
\item[Line 3:] Definition of $L({\rm data}|s)$, up to a constant.  The syntax {\tt d[[i]]} represents the element $d_i$.  The expression is not identical to eq.~\ref{eq:likelihood}; it has been manipulated algebraically to avoid the factorial term, which consumes CPU.  The manipulation starts by computing the logarithm of the likelihood:
\begin{eqnarray}
\nonumber \log L({\rm data}|s) &=& \sum_{i=1}^N \{d_i \log(b_i+s\cdot f_i)-\log(d_i!)-(b_i+s\cdot f_i)\} \\
\nonumber &=& \sum\{d_i \log(b_i+s\cdot f_i)\}-\sum \{\log(d_i!)+b_i\} - s\cdot A \\
&=& \sum\{d_i \log(b_i+s\cdot f_i)\} - s\cdot A + {\rm const.}
\end{eqnarray}
The factorial is hidden in the last term, which is constant in $s$.  
The likelihood is:
\begin{eqnarray}
\label{eq:logL}
\nonumber L({\rm data}|s) &=& \exp\{\log L({\rm data}|s)\} \\
&=& {\rm const.} \cdot \exp \left( \sum\{d_i \log(b_i+s\cdot f_i)\} - s\cdot A\right),
\end{eqnarray}
which is exactly what is written in line 3, except for the constant factor which is ignored, because it is absorbed in the normalization constant that is going to be computed in line 5.
\item[Line 4:] The prior is defined here (see Section \ref{sec:prior}).  It doesn't have to be normalized, because the posterior will be eventually normalized in one step, using the normalization constant that is going to be computed in line 5.  In this example, the prior is a unit step function, which is 0 for $s<0$ and 1 for $s \ge 0$.  This is a ``flat'' prior, except for excluding a-priori negative values for $s$, assuming that this would not make physical sense.  In cases where $s$ would make sense to be negative, it is possible to replace the above line 4 with 
\begin{lstlisting}
Prior[s_] := 1
\end{lstlisting}
However, if that is done, it will be seen in the next steps that the posterior will be impossible to normalize (it will be ``improper'').  This can be avoided by cutting off the prior at some extreme values of $s$.  For example, to define a flat prior between -2000 and 1000, one can replace line 4 with
\begin{lstlisting}
Prior[s_]:= UnitStep[s-(-2000)] * UnitStep[1000 - s]
\end{lstlisting}
When $L({\rm data}|s)$ converges to 0 quickly for large $|s|$, it doesn't matter whether the prior is cut off at $\pm$1000 or $\pm$2000 or any other large number.  The cut off is introduced for purely numerical reasons, to avoid infinities; the result does not depend practically on the exact cut off value.

Another detail is that, if $s$ can be negative, then there is a danger of some $(b_i+s\cdot f_i)$ assuming negative values, for which the likelihood is not defined, because Poisson probability is not defined with negative background.  To avoid such problems, line 3 should be re-written, inserting a {\tt Max} function that never allows any of these terms to become negative:
\begin{lstlisting}
L[s_] := Exp[Sum[d[[i]]*Log[Max[0, b[[i]] + s*f[[i]]]], {i, 1, nBins}] - s*A]
\end{lstlisting}

It is very easy to manipulate line 4 to encode any prior; not only flat priors.  Since this is possible, it is also possible to demonstrate the sensitivity of the result on the prior, by trying out various priors. If one accidentally defines a prior which makes the posterior improper, that will become obvious in the next step where the normalization constant is computed for the posterior, and one can then go back to line 4 and cut off the prior at some high value of $s$ to avoid the divergence of the normalization constant.  It is easy to try different cut off values, to confirm that the posterior does not depend on this cut off.

\item[Line 5:] By numerical integration, the normalization constant $\mathcal N$ of eq.~\ref{eq:N} is computed, and stored as {\tt NormConst}.  If the posterior is improper, this command will fail, and the measures mentioned above can be taken.

\item[Line 6:] Here the posterior is finally defined, according to eq.~\ref{eq:posteriorPoisson}.
\end{description}

\begin{figure}
\subfigure[]{\includegraphics[width=0.5\textwidth]{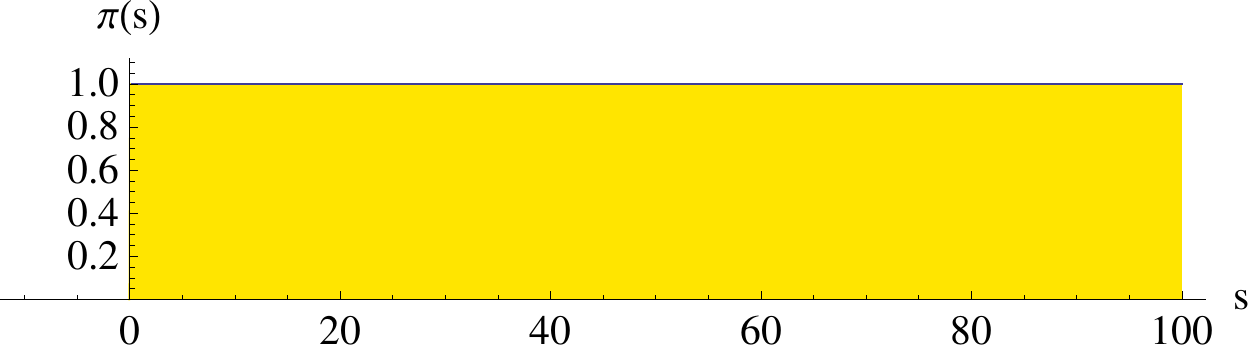}}
\subfigure[]{\includegraphics[width=0.5\textwidth]{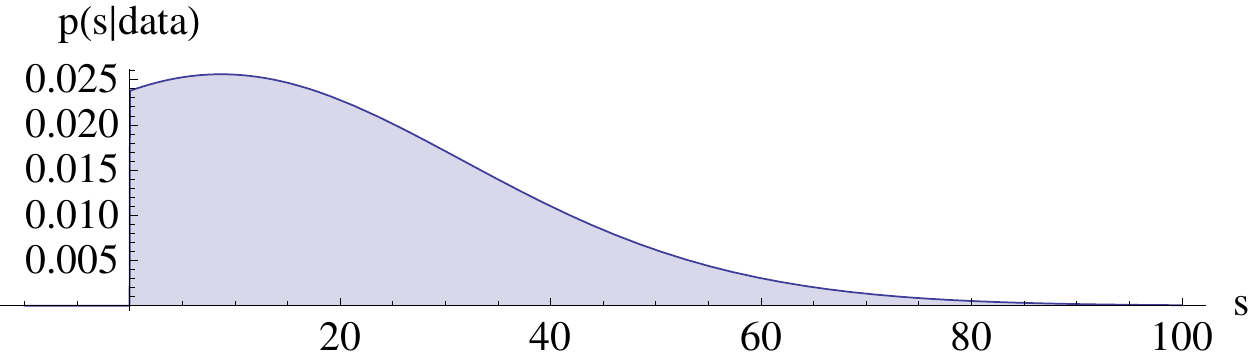}}
\subfigure[]{\includegraphics[width=0.5\textwidth]{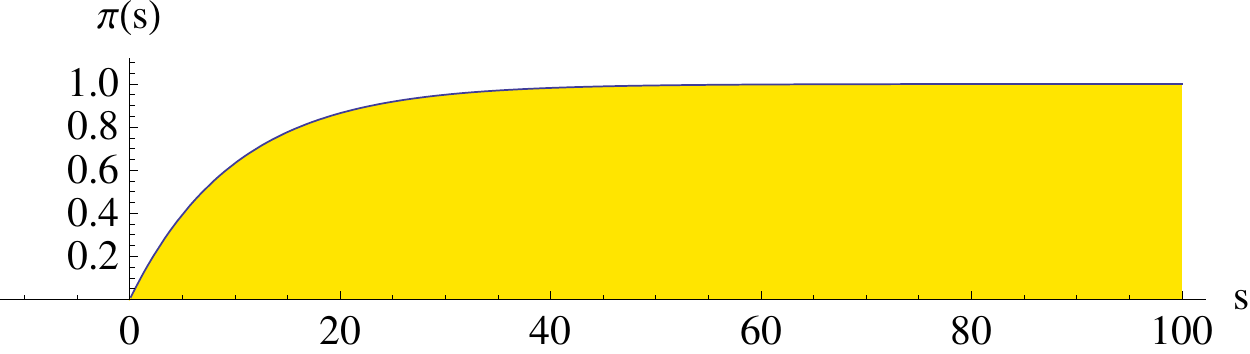}}
\subfigure[]{\includegraphics[width=0.5\textwidth]{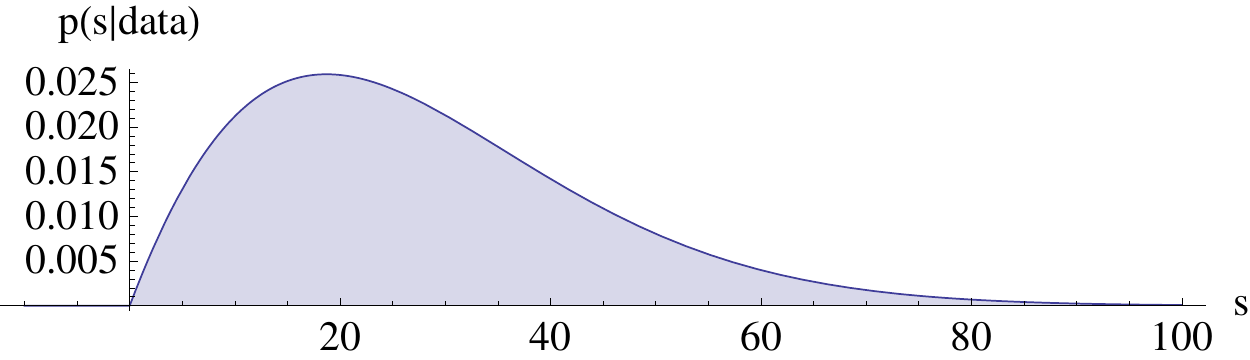}}
\subfigure[]{\includegraphics[width=0.5\textwidth]{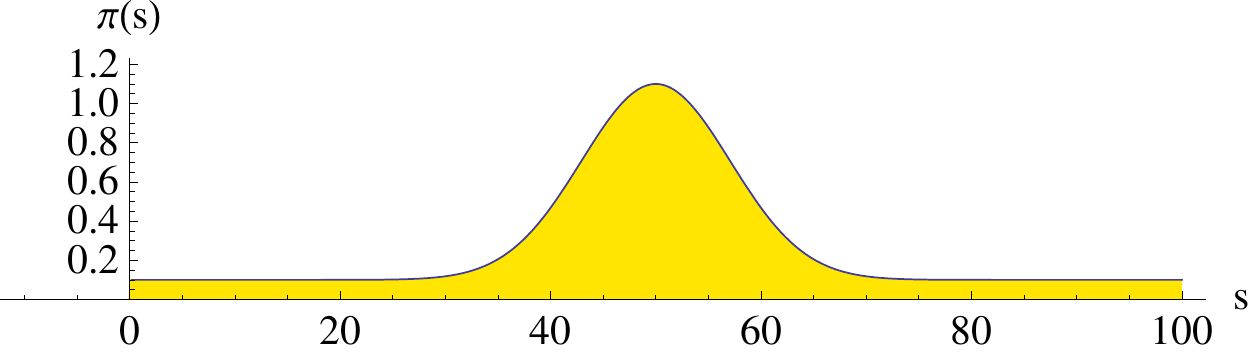}}
\subfigure[]{\includegraphics[width=0.5\textwidth]{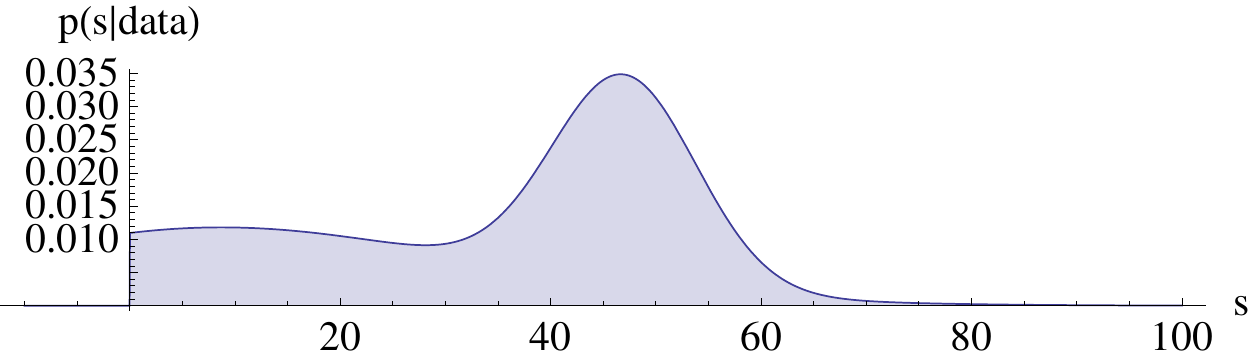}}
\caption{Three pairs of priors (not normalized) on the left and their corresponding posteriors on the right.  The quantity of interest ($s$) which is estimated, and the data on which its inference is based, are explained in Section~\ref{sec:poisson}.\label{fig:posterior1}}
\end{figure}

It is now easy to visualize the posterior and define credibility intervals from it.  For example:
\begin{lstlisting}
Plot[Posterior[s],{s, -50, 100},AxesLabel->{"s", "p(s|data)"}]
NIntegrate[Posterior[s], {s, -Infinity, Infinity}]
FindRoot[NIntegrate[Posterior[s], {s, 0, x}] - 0.95, {x, 10}]
\end{lstlisting}
Line 1 produces a plot similar to those in Fig.~\ref{fig:posterior1}, and line 2 confirms that $\int_{-\infty}^{+\infty} p(s|\text{data})ds = 1$.  Line 3 computes numerically the 95\% quantile of $p(s|\text{data})$, namely the 95\% credibility level upper limit on $s$, which is this exaple is 55.7 events.\footnote{The number 10 which appears in the command is the initial value of {\tt x} that is used to find numerically the root of the equation $\int_0^x p(s|\text{data}) ds = 0.95$.  It can be different, obviously.}

%%%%%%%%%%%%%%%%%%%%%%%%%%%%%%%%%%%%%%%%%%%%%%%%%%%%%%%%%%%%%%%%%%%%%%%%
\section{Binomial-distributed data}
\label{sec:binomial}

The measured quantity is not always a Poisson-distributed variable.  For example, consider the dijet angular distribution analysis by ATLAS \cite{ATLASdijetAndAngular}.   The observable, $F_\chi$, is the fraction of events which are central in each dijet mass bin.   The exact definition of ``central'' is of no importance here.  In general, there is some criterion, and each event represents a Bernoulli trial, leading to success if the criterion is satisfied.

The probability of having $t$ successes in $T$ trials, when each success has probability $\epsilon$, is given by the Binomial distribution:
\begin{equation}
P(t|\epsilon,T) = \binom{T}{t} \epsilon^t (1-\epsilon)^{T-t}
\end{equation}

In each bin $i$ of the total $N$ dijet mass bins, the observed number of events in bin $i$ be $T_i$, of which $t_i$ are central.  Let the probability of non-signal (i.e.\ background) events be $\epsilon_{\text{bkg},i}$. 

A theorist knows the probability of signal events to be central when they belong in bin $i$, which is denoted $\epsilon_{\text{sig},i}$.  He also knows, like in Section~\ref{sec:poisson}, the {\em total} (not only central) number of signal events in bin $i$, which is $s\cdot f_i$.

As before, $s$ is the POI.  The likelihood of the observed data, assuming some value for $s$, is

\begin{eqnarray}
L(T_i,t_i|s) &=& \prod_{i=1}^{N} \binom{T_i}{t_i} (\epsilon_i)^{t_i} (1-\epsilon_i)^{T_i - t_i}, \\
\text{where } \epsilon_i &=& \frac{1}{T_i}(\epsilon_{\text{bkg},i}(T_i-s\cdot f_i) + \epsilon_{\text{sig},i} \cdot s\cdot f_i)
\end{eqnarray}

The computation in this case is a little more complicated than the simple Poisson case, because the inputs are more.  The data are not one array, but two: $T_i$ and $t_i$.  The signal is also described by two arrays: $f_i$ and $\epsilon_{\text{sig},i}$.  

We can use, like in eq.~\ref{eq:logL}, the logarithm of $L(T_i,t_i|s)$, to simplify and speed up the computation:

\begin{eqnarray}
\nonumber \log L(T_i,t_i|s) &=& \sum_{i=1}^{N} \left\{ \log\binom{T_i}{t_i} + t_i\log\epsilon_i + (T_i-t_i)\log(1-\epsilon_i) \right\} \\
&=& \text{const.} + \sum t_i \log(\epsilon_i) + (T_i-t_i)\log(1-\epsilon_i) \Rightarrow\\
L(T_i,t_i|s) &=& \text{const.}\cdot \exp\left\{\sum t_i \log(\epsilon_i) + (T_i-t_i) \log(1-\epsilon_i)\right\} \label{eq:logLbinom}
\end{eqnarray}

Let's create an example, where for simplicity $\epsilon_{\text{sig},i}$ is constant in all bins $i$, and equal to $\epsilon_{\text{sig}} = 0.5$.  Let's use for $f_i$ the same values that we used in Section~\ref{sec:poisson}.  For the background we will assume that $\epsilon_{\text{bkg},i} = \epsilon_{\text{bkg}} = 0.1$ for all bins $i$.  We will use as $t_i$ the same array that we called $d_i$ in the example of Section~\ref{sec:poisson}, and we will add a new array $T_i$.  To make the example look like a scenario without new physics, we will define $T_i$ by sampling a Poisson distribution with mean $t_i / \epsilon_{\text{bkg}}$.  Let's put the above in code:

\begin{lstlisting}
T = {208837, 144387, 102698, 70993, 48508, 34414, 23578, 15452, 10461, 7127, 5078, 3767, 2160, 1591, 1098, 739, 437, 284, 244, 148, 13, 78, 21, 13, 7, 10, 18, 8, 5, 5};
t = {20839, 14404, 10285, 7094, 4841, 3440, 2338, 1555, 1059, 706, 515, 367, 214, 155, 112, 73, 45, 31, 23, 14, 2, 9, 2, 1, 1, 0, 2, 1, 0, 0};
nBins = Length[t];
epsilonSig = Table[0.5, {i, 1, nBins}];
epsilonBkg = Table[0.1, {i, 1, nBins}];
f = {0, 2.1*10^-05, 0.00067, 0.00097, 0.0003, 0.0016, 0.0023, 0.0085, 0.0044, 0.0068, 0.0099, 0.011, 0.019, 0.036, 0.037, 0.056, 0.17, 0.42, 0.17, 0.025, 0.0088, 0, 2.1*10^-05, 0, 0, 0.00067, 0, 0, 0, 0}/2;
\end{lstlisting}

\begin{figure}
\includegraphics[width=\textwidth]{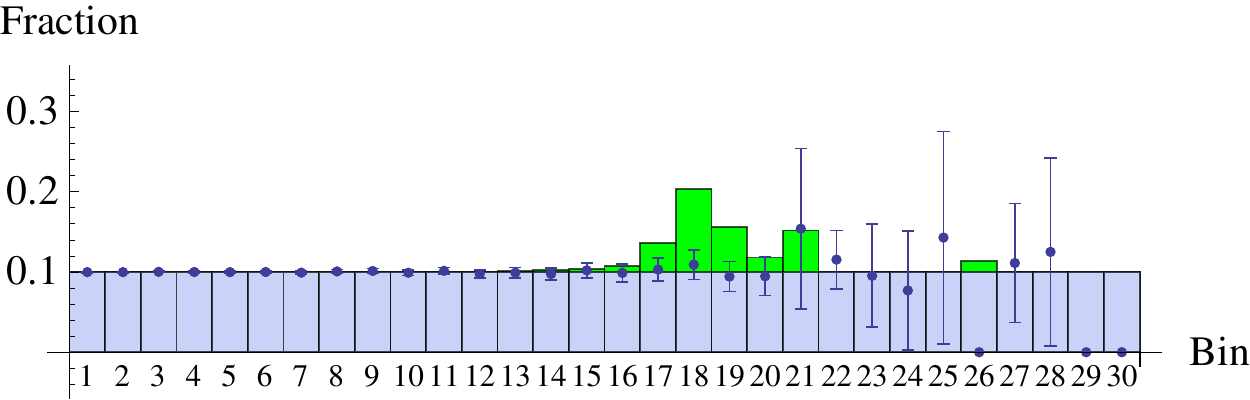}
\caption{An example of data (markers), background (blue bars), and assumed signal distribution for $s=500$ (green bars).  Details about the chosen example are given in Section~\ref{sec:binomial}. The vertical axis shows the fraction of events which satisfy a criterion, e.g.\ being central.  The error bars are, for simplicity, just the Pearson interval that spans $\pm \sqrt{\frac{t_i}{T_i}(1-\frac{t_i}{T_i})/T_i}$.  \label{fig:dataBinom}}
\end{figure}

The above choice of numbers is depicted in Fig.~\ref{fig:dataBinom}.

Now that the inputs are defined, let's write the computational part:
\begin{lstlisting}
epsilon[i_,s_]:=(epsilonBkg[[i]]*(T[[i]] - s*f[[i]]) + epsilonSig[[i]]*s*f[[i]])/T[[i]]
L[s_] := Exp[Sum[t[[i]]*Log[epsilon[i, s]] + (T[[i]] - t[[i]])*Log[1 - epsilon[i, s]], {i, 1, nBins}]]
Prior[s_] := UnitStep[s]
NormConst = NIntegrate[L[s]*Prior[s], {s, -Infinity, Infinity}]
\end{lstlisting}

Above, line 1 obviously defines $\epsilon_i$ as a function of $s$ and $i$, which is then used in line 2 that reproduces eq.~\ref{eq:logLbinom}, except for the constant term which is omitted on purpose, to be included in the overall normalization constant that is computed in line 4.  Line 3 defines the prior $\pi(s)$ of our choice (not normalized), which here is uniform in $s>0$, and line 4 computes the normalization constant $\mathcal{N} = \int L(s)\pi(s) ds$.  

At this point, in the example we use, a computational difficulty appeared.  This is an opportunity to explain how to overcome it, and how to investigate such cases.
When we executed line 4, a complain was returned that the numerical integration didn't converge, and a half-done result was returned, which was $0.*10^{276465283}$, which looked like an approximation of $0\times\infty$.  To understand what was going on, we tried to plot directly $L(s)$, using the command {\tt Plot[L[s],{s,0,100}]}, however that failed too, so, no wonder the integral was failing.  Then instead of plotting $L(s)$ we tried something more humble; to just compute it for $s=10$, and for $s=20$.  The result was two very small numbers, with a similar order of magnitude: $3.9\times 10^{-96226}$ and $4.0\times 10^{-96226}$.  This is a sign that the likelihood function is computable (it had no reason to not be anyway), but its extremely small numerical value makes its plotting and integration problematic.  Solution:  Remember that we can multiply the likelihood with any constant, since in the end the posterior it will be normalized to 1 anyway.  It would makes computation easier to divide the likelihood it by a constant of the same order of magnitude, to transform $3.9\times 10^{-96226}$ to 3.9, i.e.\ a much easier number to treat.  One way to do this is to divide $L(s)$ by $L(0)$.  This is indeed done in the following code:

\begin{lstlisting}
NormConstDividedByL0 = NIntegrate[L[s]/L[0]*Prior[s], {s, -Infinity, Infinity}]
NormConst = NormConstDividedByL0 * L[0]
Posterior[s_] := L[s]*Prior[s]/NormConst
\end{lstlisting}

The trick we played was to add the division by {\tt L[0]} in the integrand of line 1, defining this auxiliary variable {\tt NormConstDividedByL0}, and then we defined {\tt NormConst} in line 2 based on {\tt NormConstDividedByL0}.  Line 3 is nothing but the definition of the normalized posterior $p(s|\text{data})$, according to Bayes' theorem, as was done in Section~\ref{sec:poisson}.  It may seem strange that dividing and multiplying by the same number makes any difference, but in computation such things can matter.\footnote{It is not necessarily the only way to treat this case.  There may be ways, by specifying a different numerical integration method for the {\tt NIntegrate} command, to make it converge without tricks.  However, showing exactly how an effective solution was worked out is more educative and can prove useful in different cases.}

\begin{figure}
\subfigure[]{\includegraphics[width=0.5\textwidth]{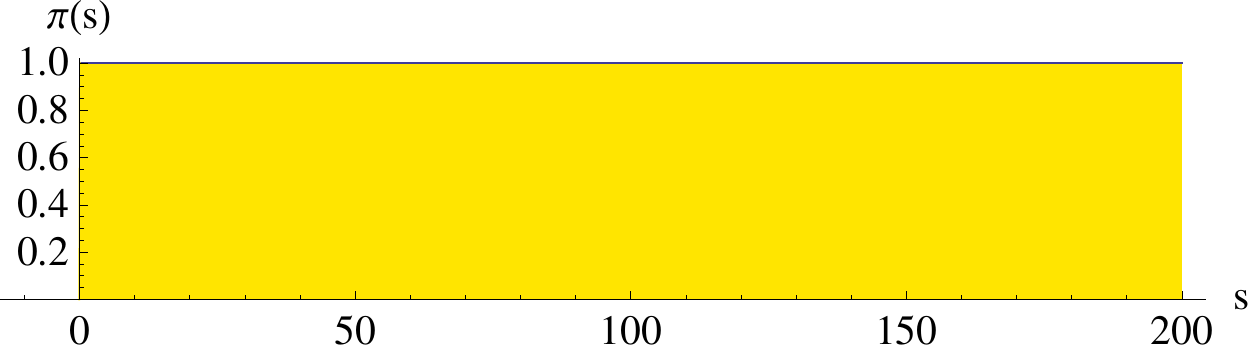}}
\subfigure[]{\includegraphics[width=0.5\textwidth]{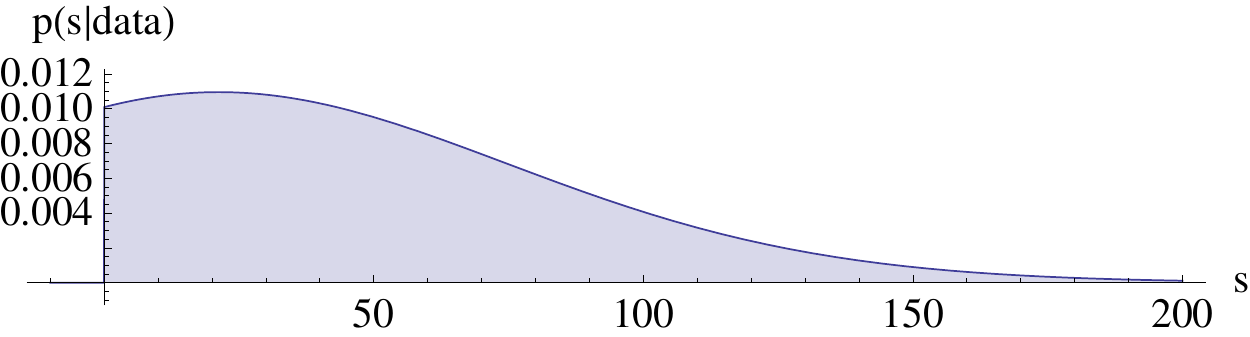}}
\subfigure[]{\includegraphics[width=0.5\textwidth]{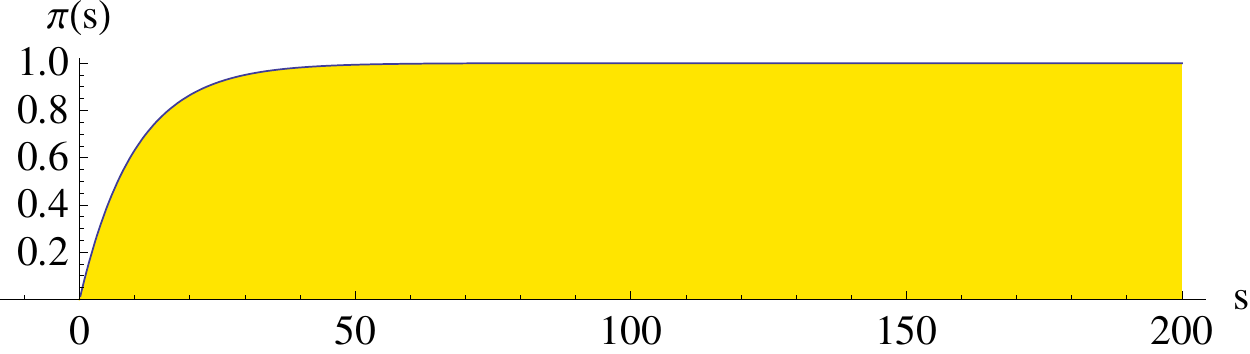}}
\subfigure[]{\includegraphics[width=0.5\textwidth]{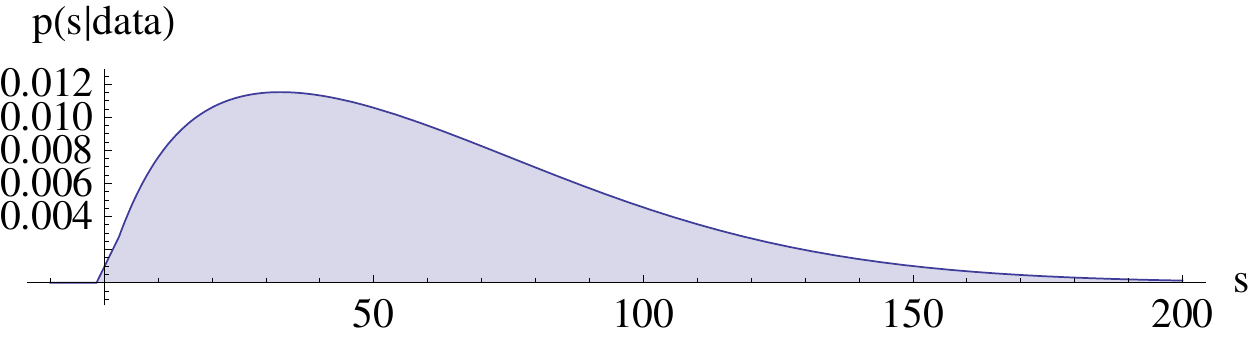}}
\subfigure[]{\includegraphics[width=0.5\textwidth]{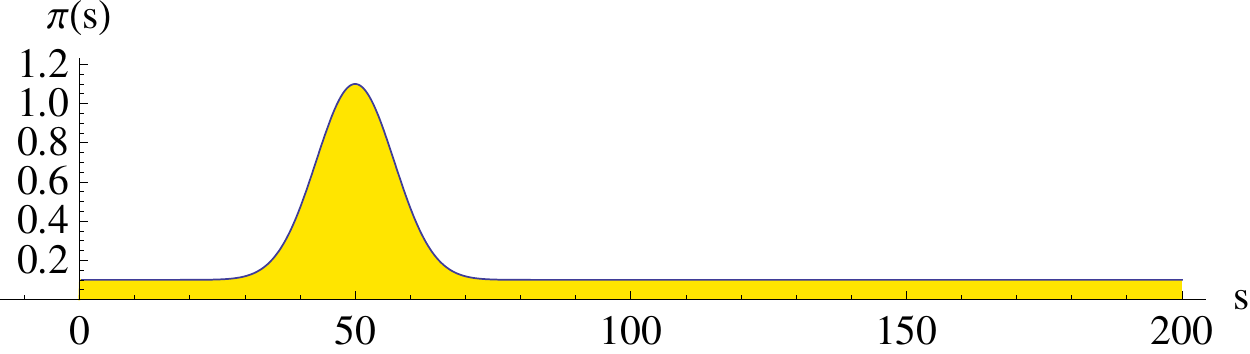}}
\subfigure[]{\includegraphics[width=0.5\textwidth]{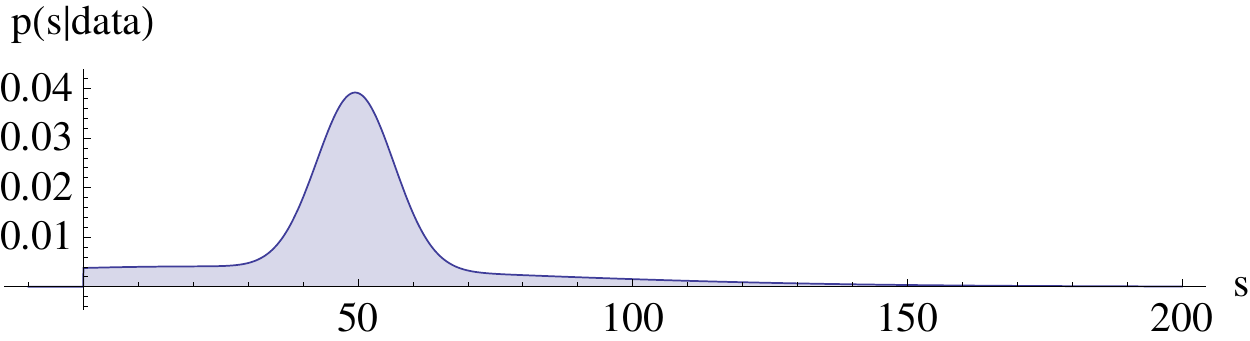}}
\caption{Three pairs of priors (not normalized) on the left and their corresponding posteriors on the right.  The quantity of interest ($s$) which is estimated, and the data on which its inference is based, are explained in Section~\ref{sec:binomial}.\label{fig:posteriorsBinom}}
\end{figure}

It is simple to plot the posterior PDF, to compute limits, and to verify that the integral of the posterior is indeed 1, using the same commands given in Section \ref{sec:poisson}.  E.g., Fig.~\ref{fig:posteriorsBinom} shows the posteriors corresponding to a variety of priors.

\section{Models where the signal is not simply additive}
\label{sec:nonAdd}

In some theoretical models, the signal is not just added to a fixed Standard Model background, but interferes with it.  As a result, the likelihood of the data, assuming some value for the POI, may not be as simple to express analytically as in eq.~\ref{eq:likelihood}.

It is still possible to set limits to such models, and to compute the posterior PDF\ of their parameter(s) of interest, as long as there is a way to map each value of the POI into a shape for the expected distribution.  This needs to be done in a continuous way, if the POI is continuous.  

To use the nomenclature of Section~\ref{sec:poisson}, one needs a function $b_i(s)$ to express the expected content of bin $i$, if the POI is $s$.  Then, the likelihood function of would in general be

\begin{equation}
L({\rm data}|s) = \prod_{i=1}^N {\rm Poisson}(d_i | b_i(s)) = \prod_{i=1}^N \frac{b_i(s)^{d_i}}{d_i!} e^{-b_i(s)}.
\end{equation}

If it is not possible to have an analytical function $b_i(s)$, one can compute the expected spectrum ($b_i$) for several discrete values of $s$, and interpolate to intermediate values of $s$ by using a {\em morphing} technique, as described in \cite{Read}.  An example of morphing would lie beyond the scope of the current document.

\section{Combining data}
\label{sec:combo}

Previously we talked about data in bins of an observable quantity.  Nothing, however, would change if the index $i$ enumerated bins of different observables, or even different experiments.  All one would do is expand the arrays to contain all independent observations.  

Combining two, or more, sets of data proceeds by writing down the joint likelihood of all observations, as a function of the POI.  In this aspect, Section~\ref{sec:poisson} \emph{was already} a combination of datasets, if we view the 30 bins as 30 independent observation, which, in that case originated from the same experiment.  We will construct also an example where we combine observations from different experiments, which is usually what people refer to by ``combination''.

Let's keep, as input from the first experiment, the numbers used in Section~\ref{sec:poisson}. We add the suffix {\tt 1} in the variable names, to remind us that they come from the first experiment.

\begin{lstlisting}
d1 = {20839, 14404, 10285, 7094, 4841, 3440, 2338, 1555, 1059, 706, 515, 367, 214, 155, 112, 73, 45, 31, 23, 14, 2, 9, 2, 1, 1, 0, 2, 1, 0, 0}; 
b1 = {21000., 14000., 10000, 7100., 4800., 3400., 2300., 1600., 1100., 740., 500, 350., 230., 160., 100, 70., 46., 30., 20., 13., 8.2, 5.2, 3.2, 2.0, 1.2, 0.71, 0.42, 0.24, 0.13, 0.074};
f1 = {0, 0.0000105, 0.000335, 0.000485, 0.00015, 0.0008, 0.00115, 0.00425, 0.0022, 0.0034, 0.00495, 0.0055, 0.0095, 0.018, 0.0185, 0.028, 0.085, 0.21, 0.085, 0.0125, 0.0044, 0, 0.0000105, 0, 0, 0.000335, 0, 0, 0, 0};
A1 = Total[f1]  (*which returns 0.494476*)
nBins1 = Length[b1]  (*returns 30*)
\end{lstlisting}

\begin{figure}
\begin{center}
\includegraphics[width=0.5\textwidth]{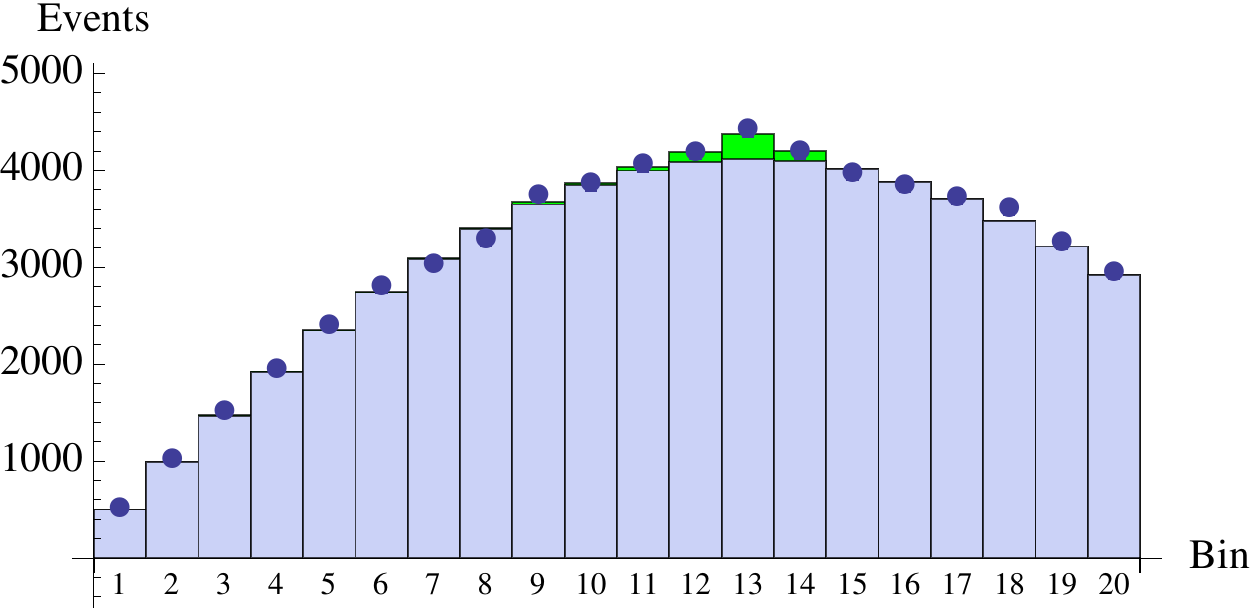}
\caption{The data (markers), expected background (blue histogram), and expected distribution of signal (green) for 600 signal events produced in the second experiment of Section~\ref{sec:combo}.\label{fig:b2andD2}}
\end{center}
\end{figure}

Let's consider a second experiment, where a different observable is used, and we have it distributed in 20 bins (instead of 30 bins that we had in the first experiment). That observable is affected by the same new physics, but the detector is different, the background level is different, the shapes of background and signal are different from the first experiment. For example,
\begin{lstlisting}
d2 = {496, 1007, 1495, 1937, 2392, 2785, 3022, 3279, 3733, 3848, 4046, 4177, 4413, 4178, 3960, 3834, 3711, 3598, 3247, 2934};
b2 = {498, 990, 1466, 1921, 2346, 2738, 3088, 3393, 3648, 3850, 3997, 4087, 4119, 4094, 4015, 3883, 3702, 3477, 3214, 2919};
f2 = {0.0016, 0.0023, 0.0085, 0.0044, 0.0068, 0.0099, 0.011, 0.019, 0.036, 0.037, 0.056, 0.17, 0.42, 0.17, 0, 0, 0, 0, 0, 0};
A2 = Total[f2]   (* which returns 0.9525 *)
nBins2 = Length[b2]  (* returns 20 *)
\end{lstlisting}
To make the example more interesting, the data of the second experiment ({\tt d2}) correspond to the background ({\tt b2}) plus 600 signal events produced in the second experiment, which are distributed according to {\tt f2}.  The elements of {\tt f2} have sum ${\tt A2} \simeq 0.95$, so, we have assumed for the second experiment most of the signal gets reconstructed.  Figure~\ref{fig:b2andD2} shows the above inputs from the second experiment.

Now that we have the data, background, and signal distribution in both experiments, we need to compute their joint likelihood, as a function of a POI, which may be some quantity proportional to the (unknown) cross-section of new physics.
For example, the POI could be the number of produced events in the first experiment, or in the second experiment, or in both experiments together, or it could be an expression of the coupling constant itself.  Let's make a choice that will spare us one proportionality constant in our expressions, and define as POI the number of signal event produced in the first experiment, which we denoted already with $s$ in Section~\ref{sec:poisson}.  Here, to remember the definition of our POI, we will denote it with {\tt s1}.  If we infer the true value of {\tt s1}, it will be easy to divide it by the integrated luminosity of the first experiment, to convert it to the cross-section of the new physics process in the conditions of the first experiment. When that is known, the coupling strength of the new physics can be extracted, which is a universal characteristic of the new physics and doesn't depend on the experiment.

The number of signal events produced in the second experiment ({\tt s2}) is proportional to the signal events produced in the first experiment ({\tt s1}).  The proportionality constant depends on the respective integrated luminosities, and on the cross-section of the new physics process in the two experiments\footnote{For example, the first experiment may be at the Tevatron, and the second at the LHC.  Different initial states, different energies, different cross-section. This difference has nothing to do with the differences between detectors, because we are talking about {\em produced} events, not reconstructed. All reconstruction effects, including detector smearing and inefficiencies, are encoded by the arrays {\tt f1} and {\tt f2}, which are independent from {\tt s1} and {\tt s2}.}.

Let's assume that the second experiment recorded 2 times larger integrated luminosity than the first experiment, and the signal cross-section in the second experiment is 3 times larger than in the first experiment.  That means that
\begin{equation}
{\tt s2} = 2\cdot3\cdot{\tt s1} \equiv r \cdot {\tt s1}
\end{equation}
We will need this proportionality constant ($r = 2\cdot3=6$) when we write the joint likelihood of the two experiments as a function of {\tt s1}.  Obviously, a theorist can calculate $r$, if he can compute the cross-section of his model in the conditions of the two experiments, and if he knows how the two integrated luminosities compare.

Time to write the joint likelihood analytically, assuming that the two experiments are statistically independent:
\begin{equation}
L({\tt d1},{\tt d2}|{\tt s1}) = \prod_{i=1}^{\tt nBins1} {\rm Poisson}({\tt d1}_i | {\tt b1}_i + {\tt s1}\cdot {\tt f1}_i) \prod_{j=1}^{\tt nBins2} {\rm Poisson}({\tt d2}_j | {\tt b2}_j + {\tt s1}\cdot r \cdot {\tt f2}_j)
\label{eq:jointL}
\end{equation}

\begin{figure}
\begin{center}
\subfigure[]{\includegraphics[width=0.45\textwidth]{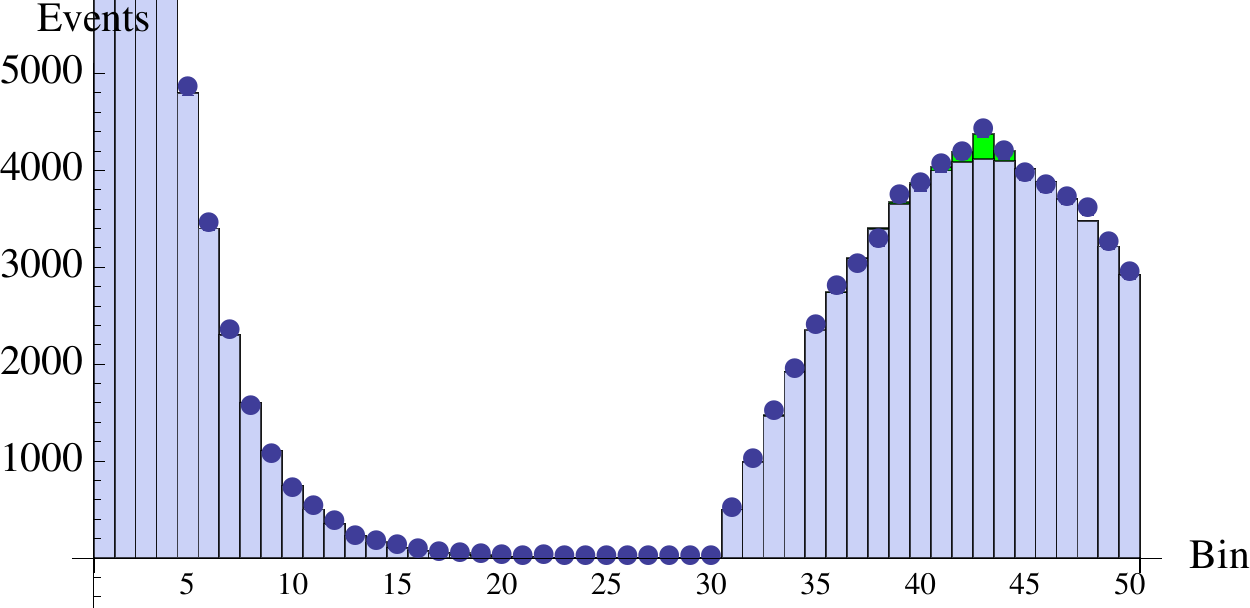}}
\subfigure[]{\includegraphics[width=0.45\textwidth]{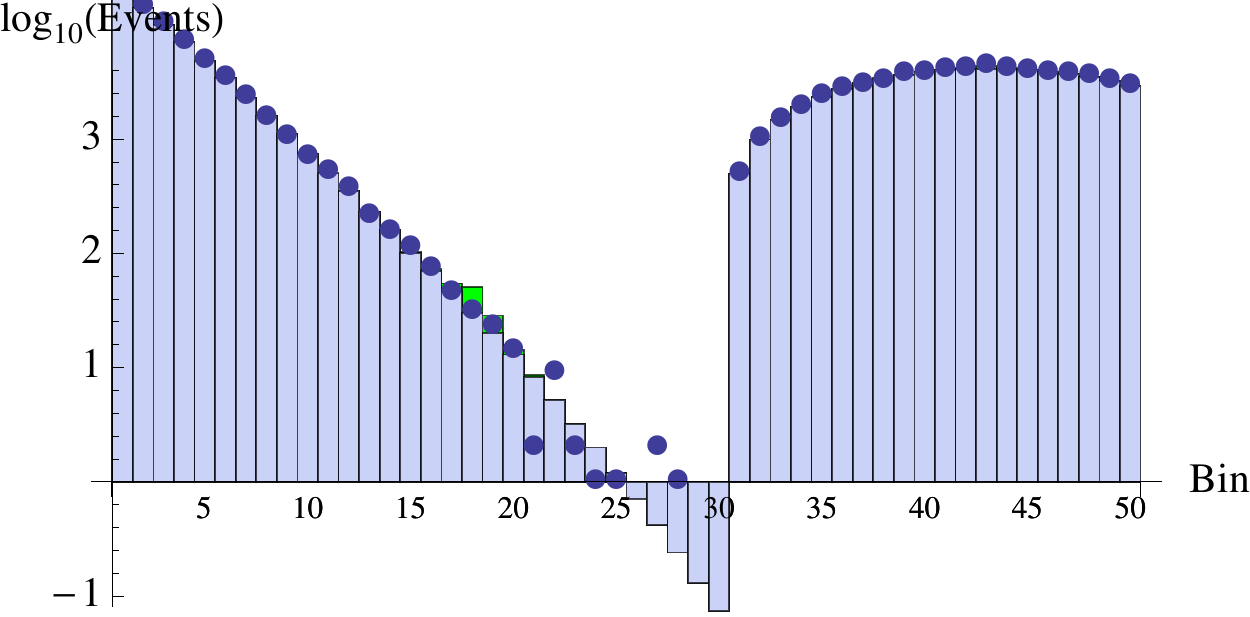}}
\caption{The data (markers), expected background (blue histogram), and expected distribution of signal (green) for 100 signal events produced in the first experiment, which correspond to 600 signal events produced in the second experiment. The plot on the left uses linear scale, which makes it easier to see the expected signal in the second experiment, and the right plot uses logarithmic scale to make visible the smaller signal expected in the first experiment. See Section~\ref{sec:combo}.\label{fig:jointData}}
\end{center}
\end{figure}

Now, let's implement this in Mathematica.  We will first merge the arrays {\tt d1} and {\tt d2} into the \emph{joint} data {\tt d}, and the arrays {\tt b1} and {\tt b2} into the \emph{joint} background {\tt b}:
\begin{lstlisting}
d = Join[d1, d2]  (* this concatenates d1 and d2 *)
b = Join[b1, b2]  (* this concatenates b1 and b2 *)
\end{lstlisting}
Then we will define a new array {\tt f} using {\tt f1} and {\tt f2}.  Note that, in eq.~\ref{eq:jointL}, all elements of {\tt f2} are multiplied by $r$.  We can simplify our expressions by defining letting {\tt f2} \emph{absorb} $r$.  This is done by writing {\tt f} as:
\begin{lstlisting}
r = 2*3    (*this is what we assume in this example*)
f = Join[f1,r*f2]  (*first f1 elements, then r*f2 elements*)
\end{lstlisting}

Figure~\ref{fig:jointData} shows the contents of {\tt b} and {\tt d}, and how the new physics would appear in this joint dataset (according to {\tt f}) if we assumed ${\tt s1} = 100$ events, i.e.\ ${\tt s2} = r\cdot s1 = 600$ events.

Using the same computational trick as in eq.~\ref{eq:logL}, we write the joint likelihood (up to a constant that will be absorbed by the normalization constant) of eq.~\ref{eq:jointL} as:
\begin{lstlisting}
L[s1_] := Exp[Sum[d1[[i]]*Log[Max[0, b1[[i]] + s1*f1[[i]]]], {i, 1, nBins1}] - s1*A1  +  Sum[d2[[i]]*Log[Max[0, b2[[i]] + s1*r*f2[[i]]]], {i, 1, nBins2}] - s1*r*A2]
\end{lstlisting}

The above expression uses explicitly the arrays of the two experiments and the constant $r$, but since we have also defined the joint arrays {\tt d}, {\tt b} and {\tt f} which absorbs $r$ in its second part, we can write the {\em totally equivalent} expression:
\begin{lstlisting}
nBins = nBins1+nBins2;  (*total bins: 30+20 = 50*)
A = Total[f];           (* sum of elements of f *)
L[s1_] := Exp[Sum[d[[i]]*Log[Max[0, b[[i]] + s1*f[[i]]]], {i, 1, nBins}] - s1*A]
\end{lstlisting}

The rest is just like before.  We define a prior PDF\ as a function of {\tt s1}, we find the normalization constant, and we get a posterior PDF:
\begin{lstlisting}
Prior[s1_] := UnitStep[s1]   (* constant for s1>0 *)
NormConst = NIntegrate[L[s1]*Prior[s1], {s1, -Infinity, Infinity}]
Posterior[s1_] := L[s1]*Prior[s1]/NormConst
\end{lstlisting}

\begin{figure}
\subfigure[]{\includegraphics[width=0.5\textwidth]{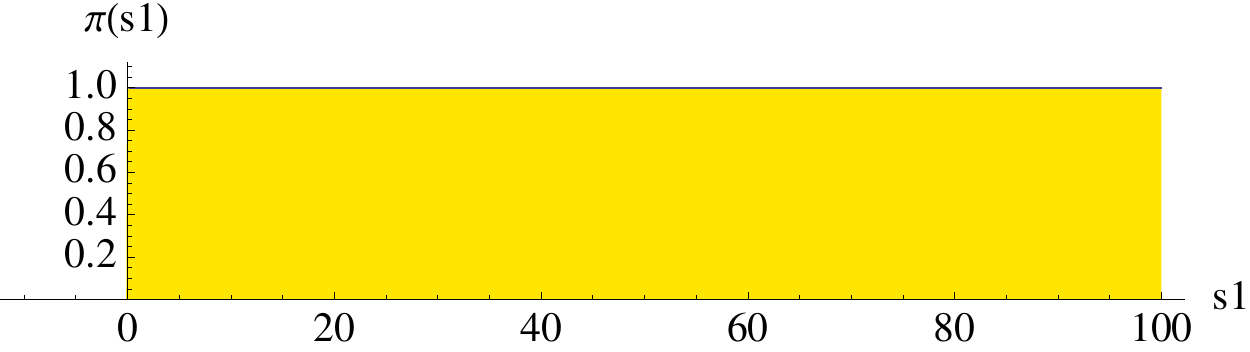}}
\subfigure[]{\includegraphics[width=0.5\textwidth]{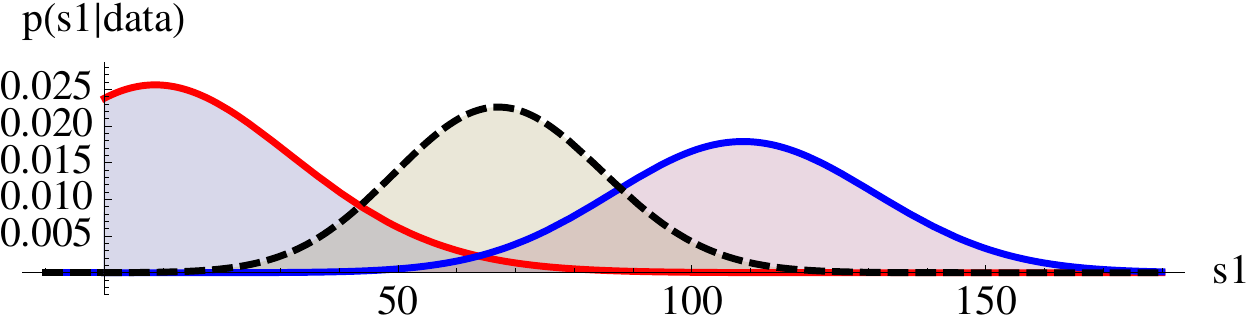}}
\subfigure[]{\includegraphics[width=0.5\textwidth]{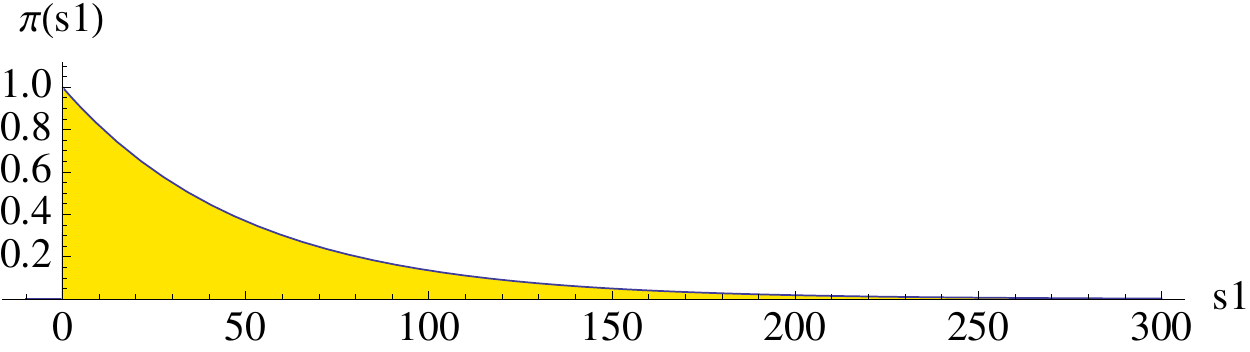}}
\subfigure[]{\includegraphics[width=0.5\textwidth]{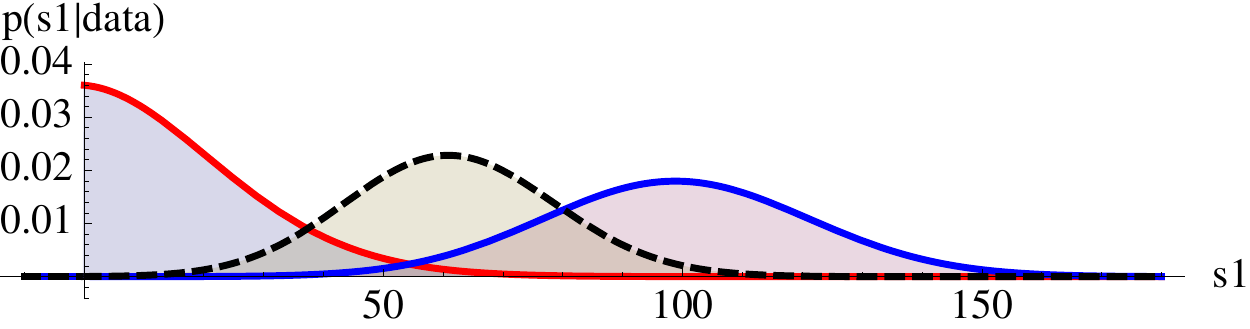}}
\subfigure[]{\includegraphics[width=0.5\textwidth]{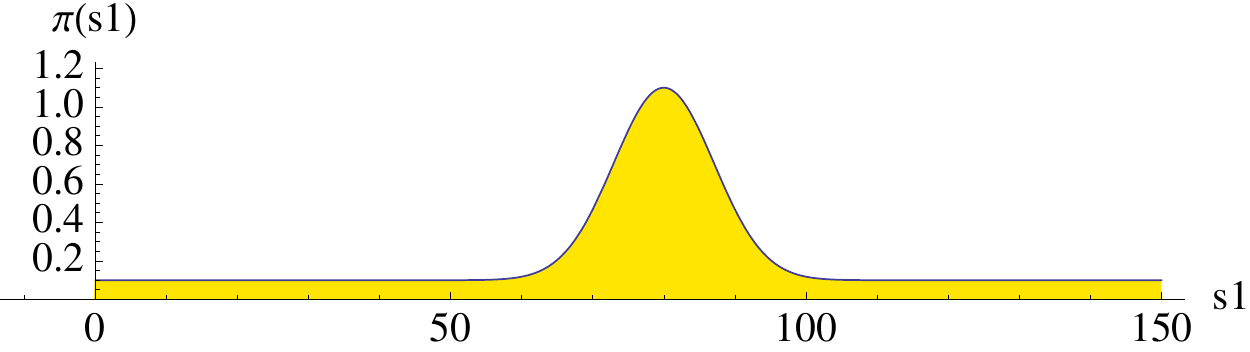}}
\subfigure[]{\includegraphics[width=0.5\textwidth]{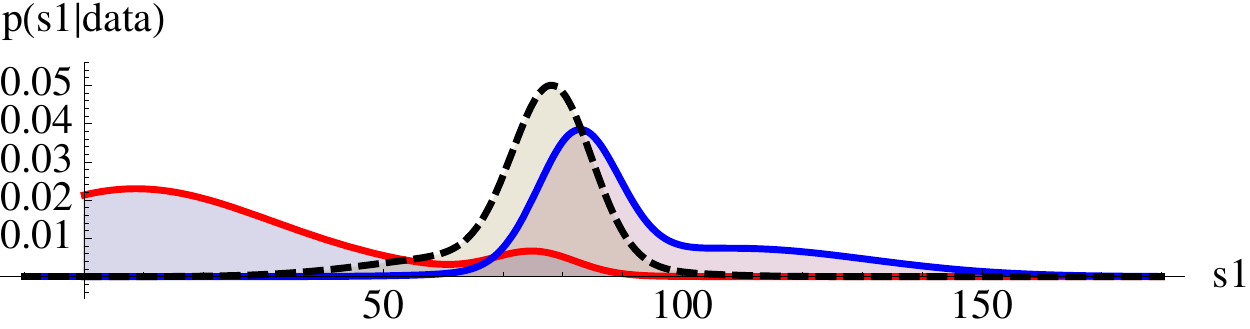}}
\caption{Three pairs of priors (not normalized) on the left and their corresponding posteriors on the right.  The red PDF\ corresponds to the posterior inferred only from the first experiment, the blue to the PDF\ only from the second experiment, and the result of the combination is shown by the black dashed PDF.  The quantity of interest ({\tt s1}) which is estimated, and the data on which its inference is based, are explained in Section~\ref{sec:combo}.\label{fig:combinationPosterior}}
\end{figure}

Figure~\ref{fig:combinationPosterior} shows the results we get from the current numerical example, with the three following indicative prior assumptions for {\tt s1}:
\begin{lstlisting}
Prior[s1_] := UnitStep[s1]
Prior[s1_] := UnitStep[s1]*(Exp[-0.02 s1])
Prior[s1_] := UnitStep[s1]*(0.1 + Exp[-(s1 - 80)^2/100])
\end{lstlisting}
Figure~\ref{fig:combinationPosterior} includes the posteriors inferred using only the first or only the second experiment.  These are computed as follows (the suffix {\tt 1} and {\tt 2} distinction the first from the second experiment):
\begin{lstlisting}
L1[s1_] := Exp[Sum[d1[[i]]*Log[Max[0, b1[[i]] + s1*f1[[i]]]], {i, 1, nBins1}] - s1*A1]
NormConst1 = NIntegrate[L1[s1]*Prior[s1], {s1, -Infinity, Infinity}]
Posterior1[s1_] := L1[s1]*Prior[s1]/NormConst1
L2[s1_] := Exp[Sum[d2[[i]]*Log[Max[0, b2[[i]] + s1*r*f2[[i]]]], {i, 1, nBins2}] - s1*r*A2]
NormConst2 = NIntegrate[L2[s1]*Prior[s1], {s1, -Infinity, Infinity}]
Posterior2[s1_] := L2[s1]*Prior[s1]/NormConst2
\end{lstlisting}

\section{Multiple parameters of interest}
\label{sec:multidim}

The POI in the above examples was always a single variable ($s$), but it can be multidimensional ($\vec{s}$).  A characteristic class of models with multiple POIs are SUSY models.  Here is an example where the data of Section.~\ref{sec:poisson} are interpreted to find a posterior in a 2-dimensional parameter space.

\begin{figure}
\begin{center}
\includegraphics[width=0.5\textwidth]{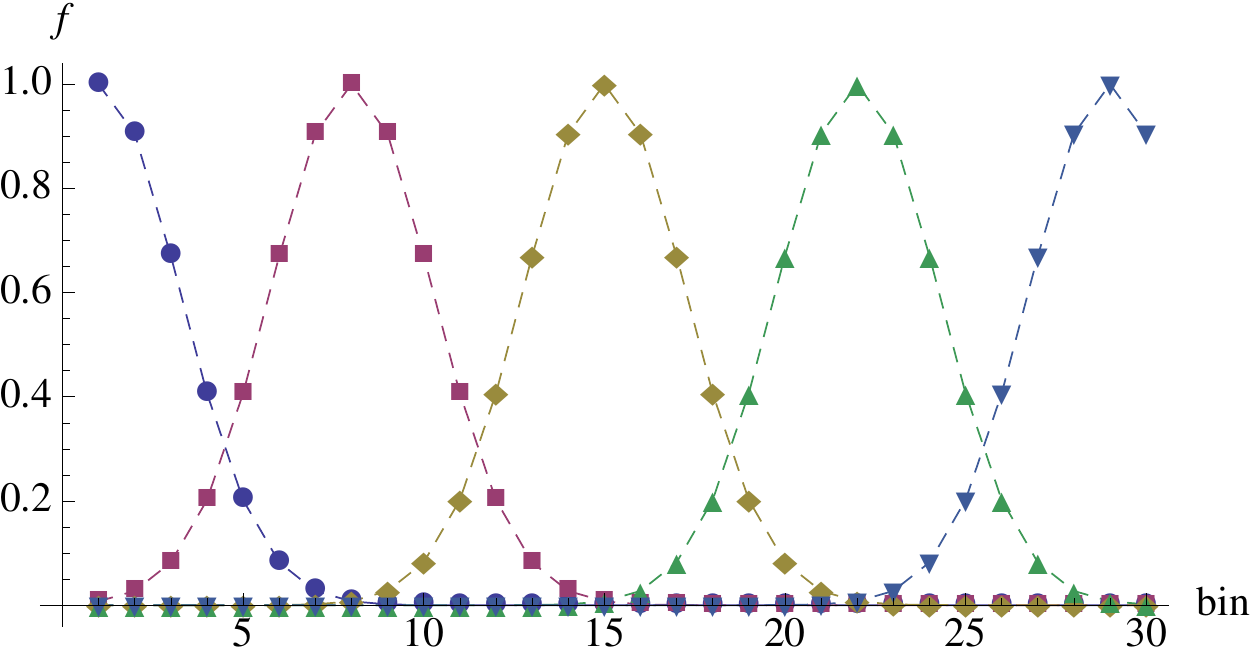}
\caption{Examples of {\tt f[s1]} defined in Section~\ref{sec:combo}, for {\tt s1} equal to 1, 8, 15, 22, and 29.\label{fig:signals}}
\end{center}
\end{figure}

Let's consider a quite generic model, where the signal is Gaussian-distributed, its mean depends on an unknown POI {\tt s1}, and its amplitude by another unknown POI {\tt s2}.  Its width could be given by a third parameter {\tt s3}, but for visualization purposes it is better to keep the space of unknown parameters 2-dimensional, so, we will assume that the width is constant.  Here is such a model:
\begin{lstlisting}
f[s1_] := Table[Exp[-(s1 - i)^2/10], {i, 1, nBins}]
\end{lstlisting}
where {\tt nBins} is the length of the {\tt b} array of Section \ref{sec:poisson}, namely ${\tt nBins} = 30$.  Figure~\ref{fig:signals} shows some examples of {\tt f[s1]} for various values of {\tt s1}.

In this example we will reuse the background array {\tt b} of Section~\ref{sec:poisson}.  
\begin{lstlisting}
b = {21000., 14000., 10000, 7100., 4800., 3400., 2300., 1600., 1100., 740., 500, 350., 230., 160., 100, 70., 46., 30., 20., 13., 8.2, 5.2, 3.2, 2.0, 1.2, 0.71, 0.42, 0.24, 0.13, 0.074};
\end{lstlisting}
To make the case more interesting, we will use data which are generated after injecting some signal on top of this background.  Specifically, the injected signal will be distributed according to $50\cdot e^{-\frac{(i-10)^2}{5}}$, where $i$ is the bin index.  This injected signal is on purpose narrower than {\tt f[10]}, to show what happens when the actual signal shape is not exactly like the hypothesis one uses to interpret the data.  So, here are the data of this example:
\begin{lstlisting}
d = {20985, 13927, 9899, 7139, 4821, 3398, 2348, 1617, 1079, 798, 555, 365, 224, 163, 88, 75, 52, 31, 21, 11, 8, 2, 5, 3, 0, 1, 0, 0, 0, 0}
\end{lstlisting}

It is simple to write the likelihood of the data, as a function of the two POIs ({\tt s1}, {\tt s2}):
\begin{lstlisting}
L0 = Exp[ Sum[d[[i]]*Log[b[[i]]], {i, 1, nBins}] ]
L[s1_,s2_] := Exp[ Sum[d[[i]]*Log[Max[0, b[[i]] + s2*f[s1][[i]]]], {i, 1, nBins}] - s2*Total[f[s1]] ] / L0
\end{lstlisting}
For computational reasons, to avoid enormous numbers, we multiply {\tt L[s1,s2]} by ${\tt L0}$, which is proportional to the likelihood of the data when no signal is assumed.  Notice that {\tt s1} is passed as an argument to {\tt f[s1]}, to determine the signal shape, and then the shape is scaled by {\tt s2}.

\begin{figure}
\begin{center}
\subfigure[]{\includegraphics[width=0.45\textwidth]{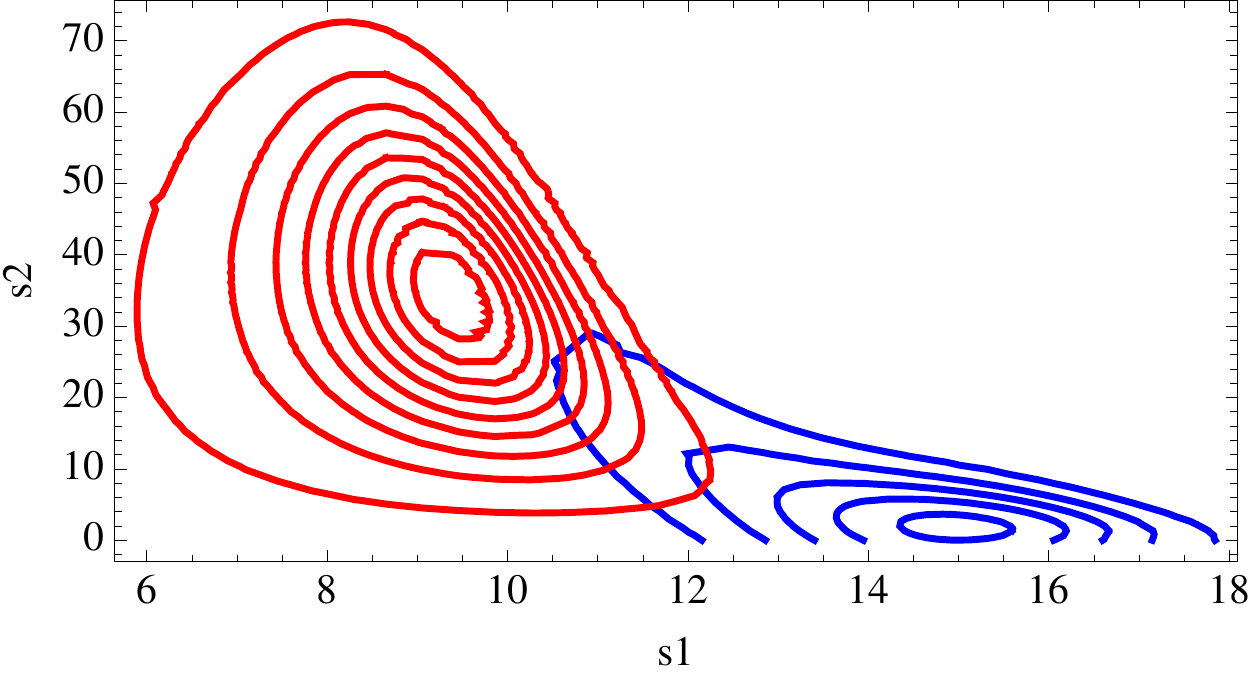} \label{fig:2dposteriors1}}
\subfigure[]{\includegraphics[width=0.45\textwidth]{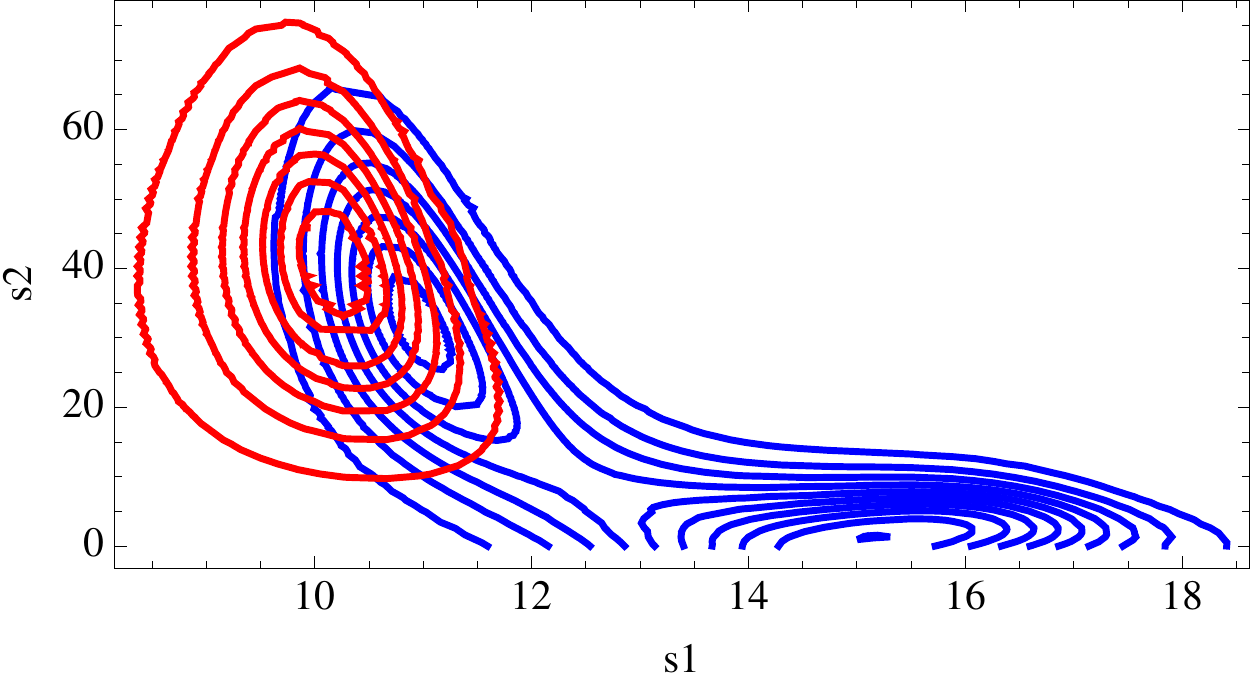} \label{fig:2dposteriors2}}
\caption{Red: Contour plot of the posterior PDF corresponding to a prior that is uniform in {\tt s2}.  Blue: The posterior corresponding to a prior where {\tt s2} is distributed around 15, according to $e^{-(i-15)^2/5}$.  Left: The signal shape ({\tt f[s1]}) assumed to compute the likelihood ({\tt L[s1,s2]}) of the data is wider than the injected signal.  Right: The ({\tt f[s1]}) has been modified to have the same width as the signal that is actually injected in the data. See discussion in Section~\ref{sec:multidim}. \label{fig:2dposteriors}}
\end{center}
\end{figure}

The prior of course needs to be defined in the same 2-dimensional space.  For example, it could represent the presumption that {\tt s2} (the produced signal amount) has to be non-negative, while all values of {\tt s1} are considered equally likely:
\begin{lstlisting}
Prior[s1_,s2_] := UnitStep[s2]
\end{lstlisting}
It is interesting to demonstrate, in the 2-dimensional case, what would happen if we introduced some non-trivial presumption in the prior.  Let's presume that {\tt s1} is more likely to be around 15 (which is at the middle of the spectrum), as expressed by the following prior:
\begin{lstlisting}
Prior[s1_, s2_] := UnitStep[s2]*Exp[-(s1-15)^2/5]
\end{lstlisting}

Figure~\ref{fig:2dposteriors1} shows the shape of the posterior (ignoring the normalization constant) for both priors.  The posterior, up to a normalization constant, is:
\begin{lstlisting}
Posterior[s1_, s2_] := L[s1,s2]*Prior[s1,s2]
\end{lstlisting}

The posterior with uniform prior, which has the same shape as the likelihood function, does not have its maximum exactly at {\tt (s1,s2)=(10,50)}, and the reason is dual: 
\begin{itemize}
\item The data ({\tt d}) are {\em consistent} with injected signal of {\tt (s1,s2)=(10,50)}, but it is ultimately the result of Poisson random fluctuations in each bin, so, it is expectable that the best-fitting {\tt (s1,s2)} will be close to that point, but not exactly there.
\item The signal shape {\tt f[s1]} that is used to compute the likelihood is wider than the actual signal that has been injected, on purpose, to demonstrate this scenario, which is quite plausible, because Nature may produce some signal, which we ignore, so we may try to interpret the data to infer the parameters of a different signal.
\end{itemize}

For comparison, Fig.~\ref{fig:2dposteriors2} shows the same result, with exactly the same {\tt b} and {\tt d}, when {\tt f[s1]} has been modified to have the same width as the injected signal:
\begin{lstlisting}
f[s1_] := Table[Exp[-(s1-i)^2 / 5], {i, 1, nBins}]
\end{lstlisting}
The difference is that, when the prior is uniform (red contours in Fig.~\ref{fig:2dposteriors2}), the posterior is more narrow in {\tt s1}.  This makes sense; it's more clear where the signal is, when we have gotten the signal width right.  As a result, the effect of the non-uniform prior is quite different in Fig.~\ref{fig:2dposteriors2} than in \ref{fig:2dposteriors1}:  The prior ``pulls'' the posterior towards {\tt s1}=15, but the likelihood is larger around ${\tt s1}\simeq 10$, so, the resulting posterior has two local maxima, of which the one near {\tt s1}=15 prevails with greater probability density.

It is also worth noting that, in both Fig.~\ref{fig:2dposteriors1} and \ref{fig:2dposteriors2}, the non-uniform prior in {\tt s1} does not only pull {\tt s1} towards 15, but it also changes the most likely value of {\tt s2}.  This happens because {\tt s1} and {\tt s2} are correlated, as one can see from the asymmetric shape of the contours.  This can only be appreciated in a multidimensional space, where there is room for correlations: The prior may be factorized to one part that depends only on {\tt s1} and another that depends only on {\tt s2}, but its effect on the posterior is not factorized in a similar way; a change in the prior with respect to {\tt s1} will modify the posterior in all dimensions.

\section{Inclusion of systematic uncertainty}
\label{sec:systematics}

Systematic uncertainties are uncertainties about assumptions which affect the measurement.  If these assumptions were slightly different, within their own (systematic uncertainty), that would have an effect on the measurement.  To quantify this effect, we need first to use parameters to quantify the assumptions.  These parameters are called ``nuisance parameters''.

The procedure to take these uncertainties into account starts by treating the nuisance parameters as if they were POIs, alongside with the actual POIs.  This leads to a multi-dimensional space of parameters, where a prior needs to be defined, and a posterior is computed based on the data.  The posterior PDF\ can be integrated along the dimension of the nuisance parameter(s), leaving only the actual POIs as free variables in the posterior.  

Let's write this analytically, denoting the nuisance parameter(s) with $n$, and the actual POI with $s$.  From Bayes' theorem,
\begin{equation}
  p(s,n|{\rm data}) = \frac{L({\rm data}|s,n) \pi(s,n) }{ \mathcal{N}} ,
\end{equation}
where 
\begin{equation}
\mathcal{N} \equiv \iint L({\rm data}|s,n)\pi(s,n)\;ds\;dn.
\end{equation}
Then, since we are not interested in the actual value of $n$, but only of $s$, the posterior we actually care about is
 \begin{equation}
p(s|{\rm data}) = \int p(s,n|{\rm data})\;dn = \frac{1}{{\mathcal{N}}} \int L({\rm data}|s,n) \pi(s,n)\;dn. \label{eq:posteriorNuis1}
\end{equation}
If the prior $\pi(s,n)$ is factorable as:
\begin{equation}
\pi(s,n) = \pi_s(s) \; \pi_n(n), \label{eq:priorFactor}
\end{equation}
then
\begin{equation}
p(s|{\rm data}) = \frac{\pi_s(s)}{\mathcal{N}} \int L({\rm data}|s,n) \pi_n(n)\;dn  
 \end{equation}
This integral can be read as ``the expected likelihood function, over all possible values of the nuisance parameter $n$'', which can be denoted:
\begin{equation}
p(s|{\rm data}) = \frac{\pi_s(s)}{\mathcal{N}} \langle L({\rm data}|s,n) \rangle_{n}  \label{eq:posteriorNuis2}
\end{equation}

Notice the similarity between eq.~\ref{eq:posteriorNuis2} and eq.~\ref{eq:posteriorPoisson}.  The only difference is that the likelihood at $s$ is replaced by the average likelihood.   If one wishes to try a different prior for $s$ he can do it by just changing $\pi_s(s)$, without having to recalculate the average likelihood.  This can be a great advantage in practical applications, where calculating the average likelihood (namely, performing the ``convolution'' of the nuisance parameters) is time-consuming.

Equation~\ref{eq:posteriorNuis2} is based on the condition that the prior is factorable as in eq.~\ref{eq:priorFactor}.  This condition is easy to satisfy, and actually most intuitive prior choices would satisfy it.  Usually the nuisance parameters express some uncertainty about the experimental conditions, like the actual detector response etc.  There is no reason to correlate, in the prior, the true cross-section of a process with the nuisances of the detector\footnote{The posterior $p(s,n|{\rm data})$ may indicate that $s$ and $n$ are correlated, but don't confuse the prior with the posterior;  eq.~\ref{eq:priorFactor} concerns just the prior.}.  However, this may not be the case for theoretical nuisance parameters, which may be intimately related to $s$ even a-priori.  After all, eq.~\ref{eq:priorFactor} refers to the prior, so, someone may wish to assume some $\pi(s,n)$ that isn't factorable, just because that's what he finds interesting.  We can not prevent that, so, the numerical examples below do not rely on the assumption of eq.~\ref{eq:priorFactor}, but make use of eq.~\ref{eq:posteriorNuis1}, which is generally true.

\begin{figure}
\begin{center}
\subfigure[]{\includegraphics[width=0.45\textwidth]{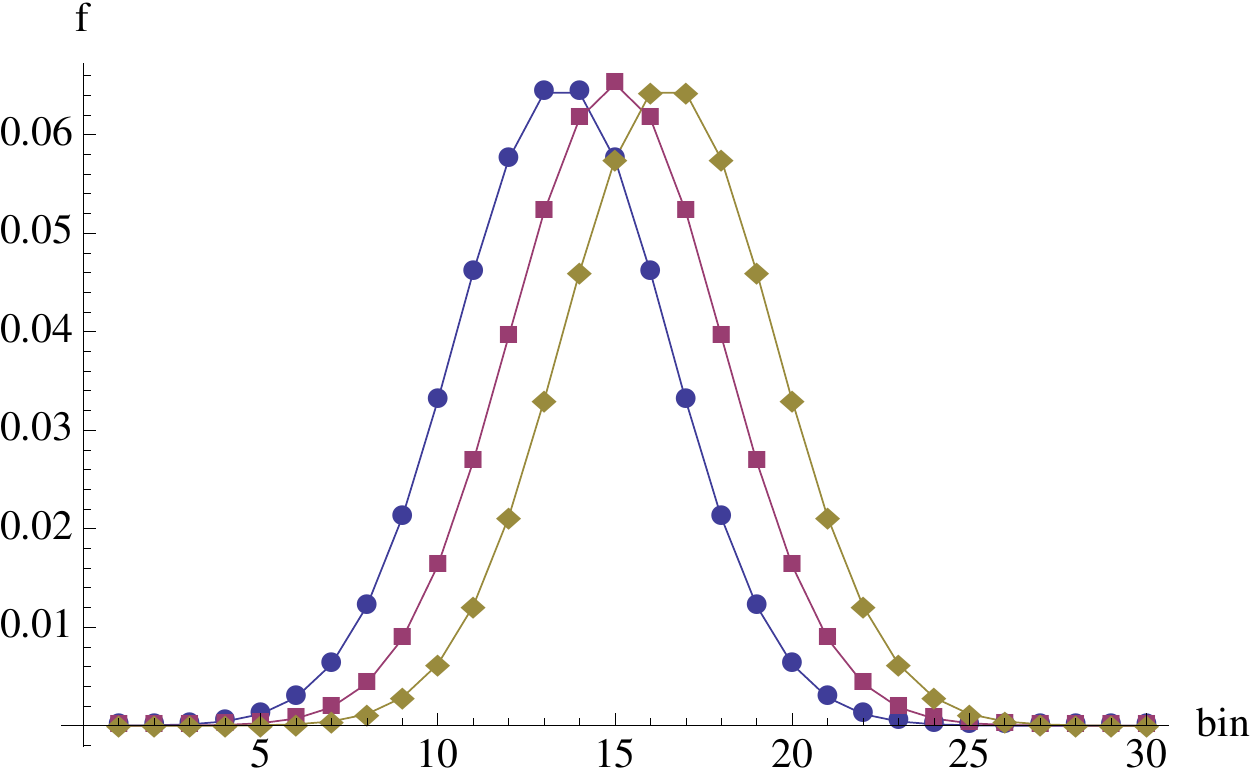} \label{fig:convSig1}}
\subfigure[]{\includegraphics[width=0.45\textwidth]{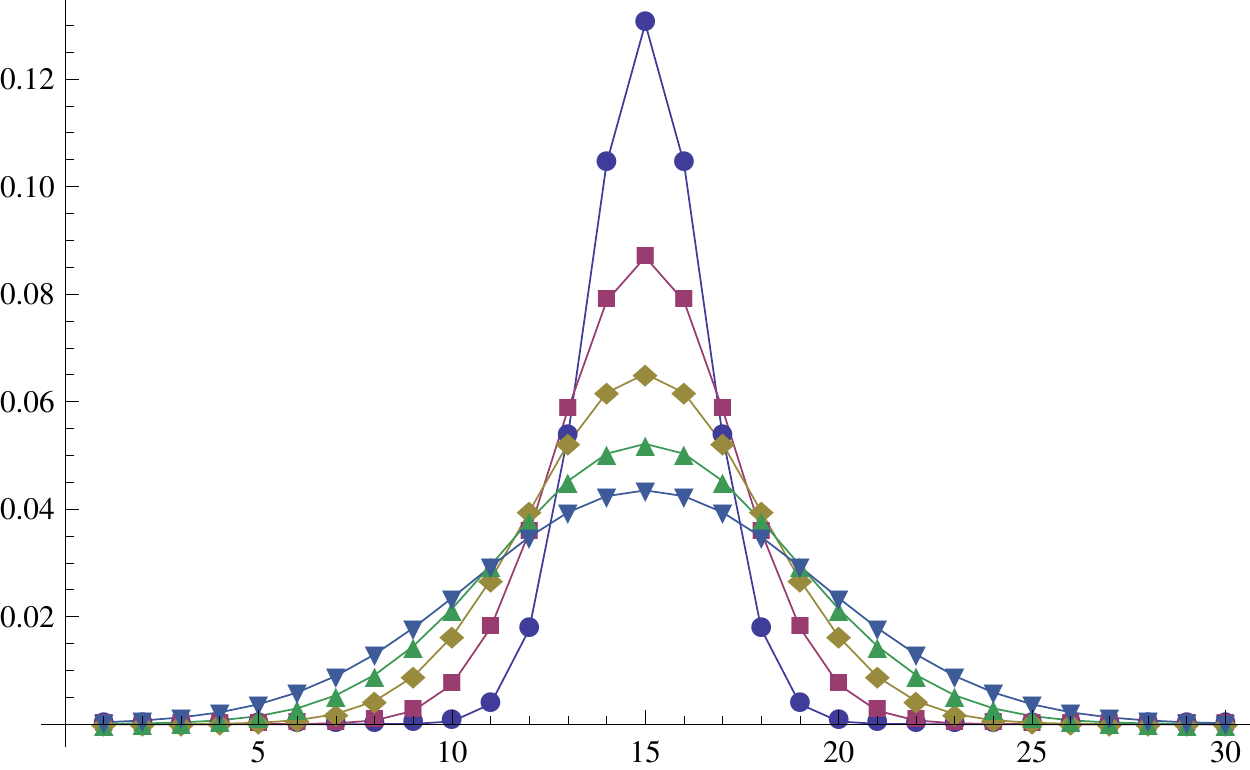} \label{fig:convSig2}}
\caption{Left: Examples of the signal {\tt f[n]} defind in Section~\ref{sec:convMean}, for {\tt n} equal to 0.9, 1 and 1.1.  
Right: Examples of the {\tt f[n]} of Section~\ref{sec:convRMS}, for {\tt n} from 0.5 to 1.5. \label{fig:convSig}}
\end{center}
\end{figure}

\subsection{Example of uncertainty in signal position}
\label{sec:convMean}

For example, let's take the background and data of the example in Section~\ref{sec:poisson}.  The goal now is to infer the amount of produced signal ($s$), where the signal is known to be Gaussian-distributed around bin 15, with standard deviation equal to 3 bins.  However, there is some doubt about the actual position of the signal; maybe the mean is not exactly 15.  This could reflect, for example, an uncertainty about the {\em actual} detector energy response, if the bins are defined in an observable which depends on energy.

Let's parametrize this uncertainty using a nuisance parameter $n$, such that the signal peaks at $15\cdot n$.  The following lines implement this parametrization.  The array {\tt f} is a function of {\tt n}, and is normalized to sum {\tt A = 0.49}, simply to keep the same acceptance as in Section~\ref{sec:poisson}.
\begin{lstlisting}
A = 0.49   (* some arbitrary acceptance *)
f[n_] := A * Table[Exp[-(15*n - i)^2/(2*3^2)], {i, 1, nBins}] / Sum[Exp[-(15*n - i)^2/(2*3^2)], {i, 1, nBins}]]
\end{lstlisting}
Figure~\ref{fig:convSig1} demonstrates this {\tt f[n]}.

Then we compute, up to a constant term which will be absorbed in the final normalization, the likelihood function {\tt L[s,n]}, which corresponds to $L(\text{data}|s,n)$ of eq.~\ref{eq:posteriorNuis1}.  To avoid problematically large numbers, we divide by the constant {\tt L0}, which is assuming no signal:
\begin{lstlisting}
L0 = Exp[Sum[d[[i]]*Log[b[[i]]], {i, 1, nBins}]];
L[s_, n_] := Exp[Sum[d[[i]]*Log[Max[0, b[[i]] + s*f[n][[i]]]], {i, 1, nBins}] - s*A]/L0
\end{lstlisting}
Note that {\tt f[n]} in this example is constructed to have  $\sum_{i=1}^{\tt nBins} f_i(n) = {\tt A} = 0.49$, for any choice of $n$.  In a more general case, where the acceptance depends on $n$, one would replace {\tt A} by {\tt Total[f[n]]}, to compute the acceptance at the same time with {\tt L[s,n]}.  This would make computation slightly slower, which is why it is avoided here.\footnote{If one uses compiled code for these computations, he will probably not notice any difference in performance, but Mathematica, in its simplest version, is an ``interpreted'' language, not compiled, which makes it considerably slower.}

\begin{figure}
\begin{center}
\subfigure[]{\includegraphics[width=0.45\textwidth]{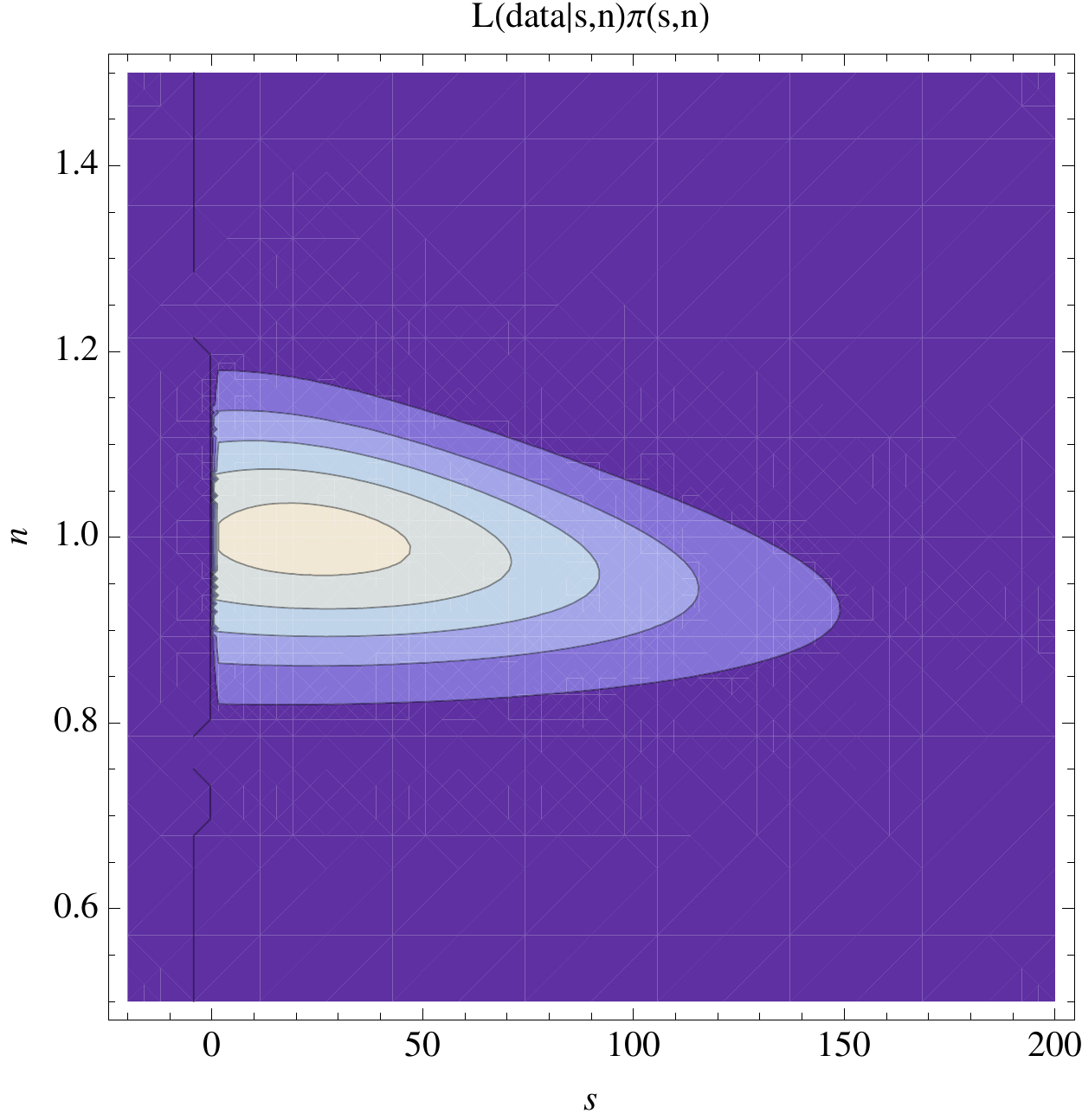} \label{fig:convPosteriorMeanSN}}
\subfigure[]{\includegraphics[width=0.45\textwidth]{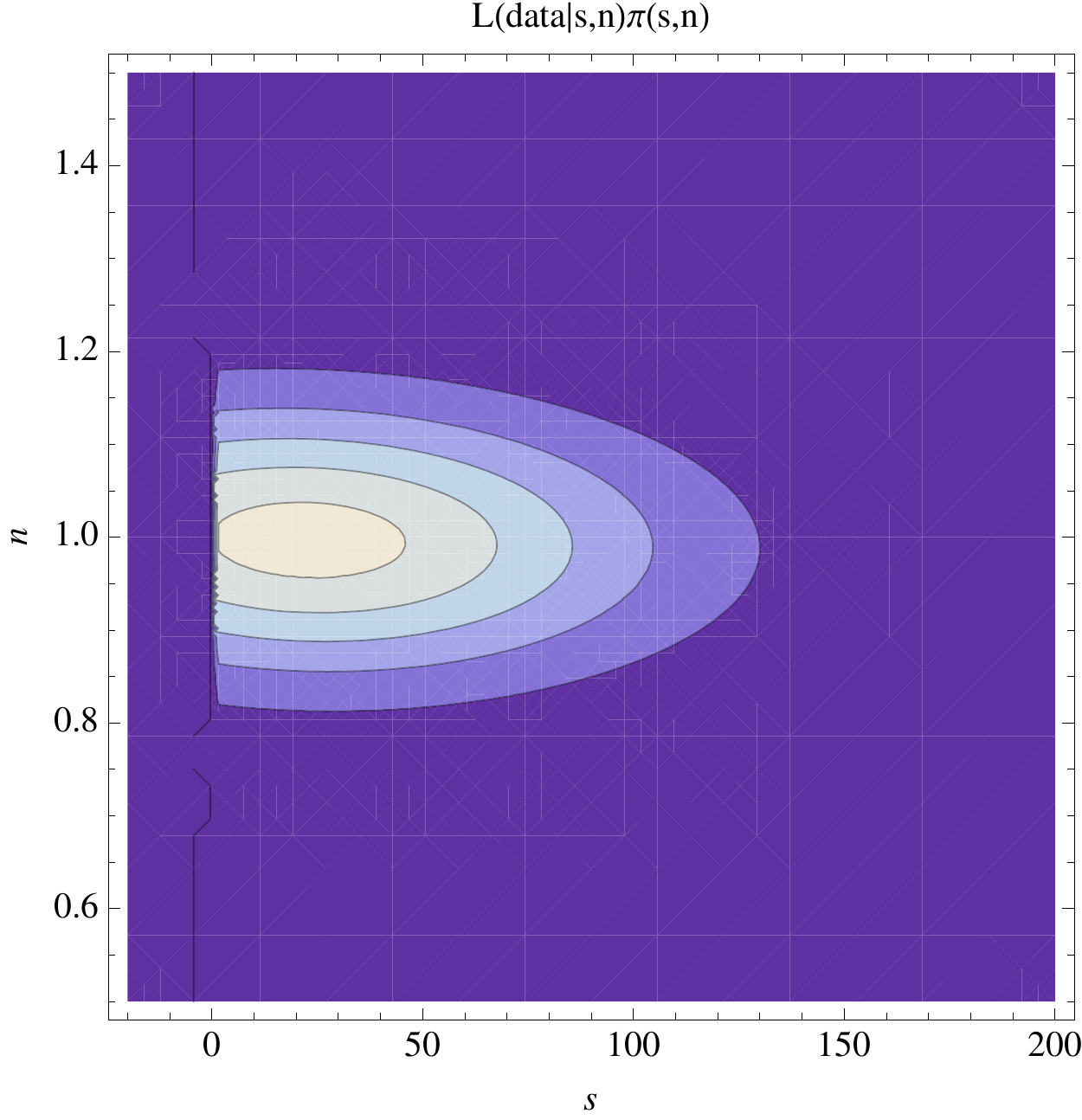} \label{fig:convPosteriorRMSSN}}
\caption{Left: The contours of $p(s,n|\text{data})$ from the example in Section~\ref{sec:convMean}.
Right: The contours of $p(s,n|\text{data})$ from the example in Section~\ref{sec:convRMS}.}
\end{center}
\end{figure}

We need to define a prior PDF (up to a constant), which will be assumed to be uniform in {\tt s}, allowing only positive values of {\tt s}, and Gaussian in {\tt n}, with maximum probability density at {\tt n = 1} and standard deviation equal to 0.1:
\begin{lstlisting}
Prior[s_, n_] := UnitStep[s] * Exp[-(n - 1)^2/(2*0.1^2)]
\end{lstlisting}

The posterior $p(s,n|\text{data})$, before integration along $n$, is given (up to a constant) by the product {\tt L[s,n] * Prior[s,n]}, which is shown in Fig.~\ref{fig:convPosteriorMeanSN}.

The next step is to ``integrate out'' $n$, in which we are not really interested.  Here, this is done using a simple approximation of the integral, where we break the interval $n\in[0.5,1.5]$ in 100 steps of size 0.01, and we approximate 
\begin{eqnarray}
\int_{-\infty}^\infty L(\text{data}|s,n) \pi(s,n)\; dn &\simeq& \int_{0.5}^{1.5} L(\text{data}|s,n) \pi(s,n) \;dn \\
&\simeq& \sum_{i=1}^{100} L(\text{data}|s, n_i) \pi(s, n_i) \cdot 0.01, \label{eq:trapezoid} \\
\text{where } n_i &=& 0.5 + i\cdot 0.01.
\end{eqnarray}
This approximation is justified by $\pi(s,n)$ being almost zero for $n>1.5$ or $n<0.5$, and eq.~\ref{eq:trapezoid} being a simple numerical integration method, admittedly not the most advanced, but simple enough to implement in the following few lines:
\begin{lstlisting}
integ[s_] := Sum[L[s, n]*Prior[s, n], {n,0.5,1.5,0.01}]
normConst = NIntegrate[integ[s], {s, -Infinity, Infinity}]
Posterior[s_] := integral[s] / normConst
\end{lstlisting}
The function {\tt integ[s]} corresponds to the result of  eq.~\ref{eq:trapezoid}.  The constant 0.01 has been omitted, or rather absorbed by the {\tt normConst} normalization constant computed in line 2.
Finally, the (normalized) posterior PDF\ of $s$ is given in line 3, which can now be plotted, or used to compute its 95\% or any other quantile, as shown in Section~\ref{sec:poisson}. 

Figure~\ref{fig:posteriorMeanS} shows the resulting posterior, after this convolution of the nuisance parameter $s$, and compares it to the posterior one would get if there were no uncertainty in $n$, namely, if $\pi(s,n)$ were  $$\pi(s,n)=\Theta(s)\cdot \delta(n-1),$$ where $\Theta(s)$ is the step function represented in Mathematica by {\tt UnitStep[s]}, and $\delta(n-1)$ is just the Kronecker $\delta$, pinning $n$ to 1.  The latter posterior, which is unaffected by systematic uncertainty, is given simply by:
\begin{lstlisting}
normConst = NIntegrate[ L[s,1]*Prior[s,1] , {s,-Infinity,Infinity}]
Posterior[s_] := L[s,1]*Prior[s,1] / normConst
\end{lstlisting}
This comparison shows that the convolution of $n$ makes the posterior wider, and the upper limit worse (looser).  Specifically, the upper limit, and 95\% credibility level, without systematic uncertainty is 131.315, and with this uncertainty it becomes 153.  However, it is not always true that inclusion of systematic uncertainty loosens the upper limit.  We will see in Section~\ref{sec:convRMS} such an example.

\begin{figure}
\begin{center}
\subfigure[]{\includegraphics[width=0.45\textwidth]{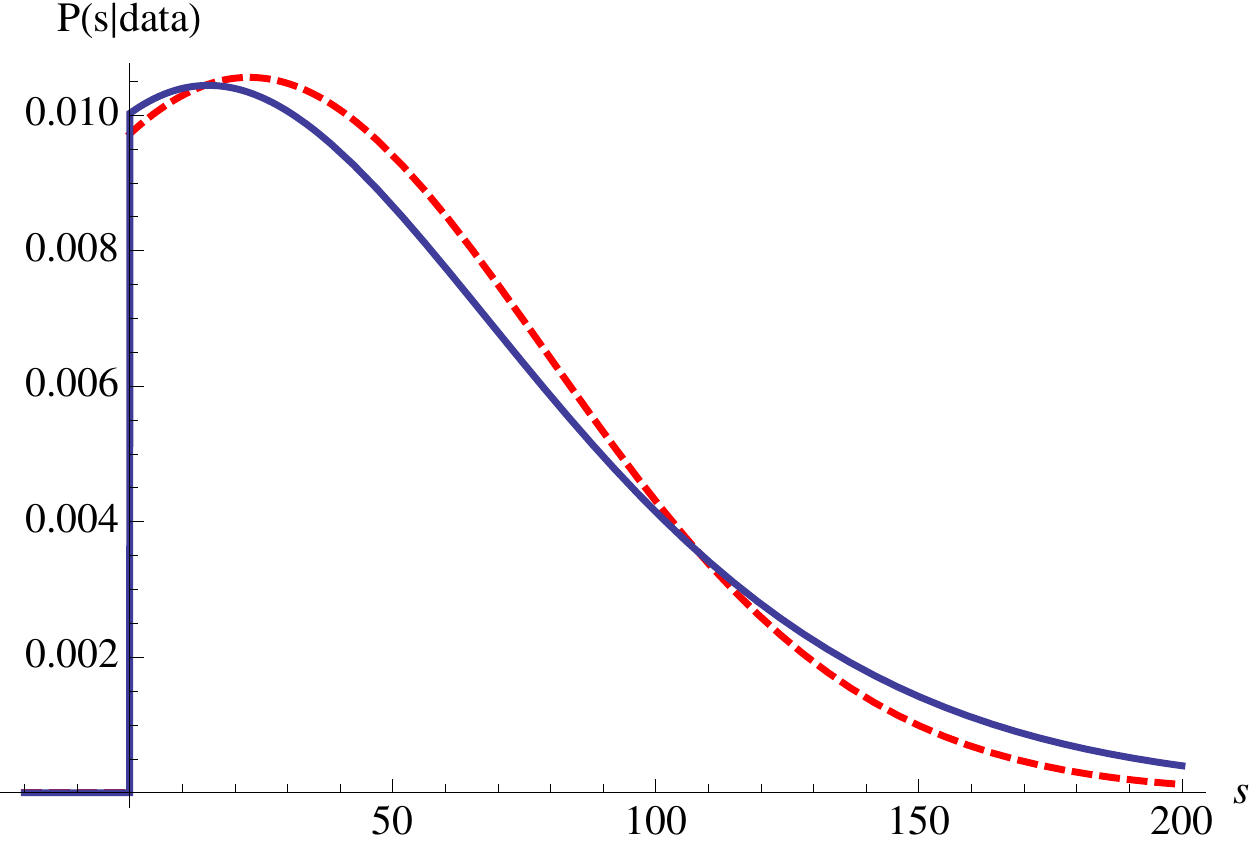} \label{fig:convPosteriorMeanS}}
\subfigure[]{\includegraphics[width=0.45\textwidth]{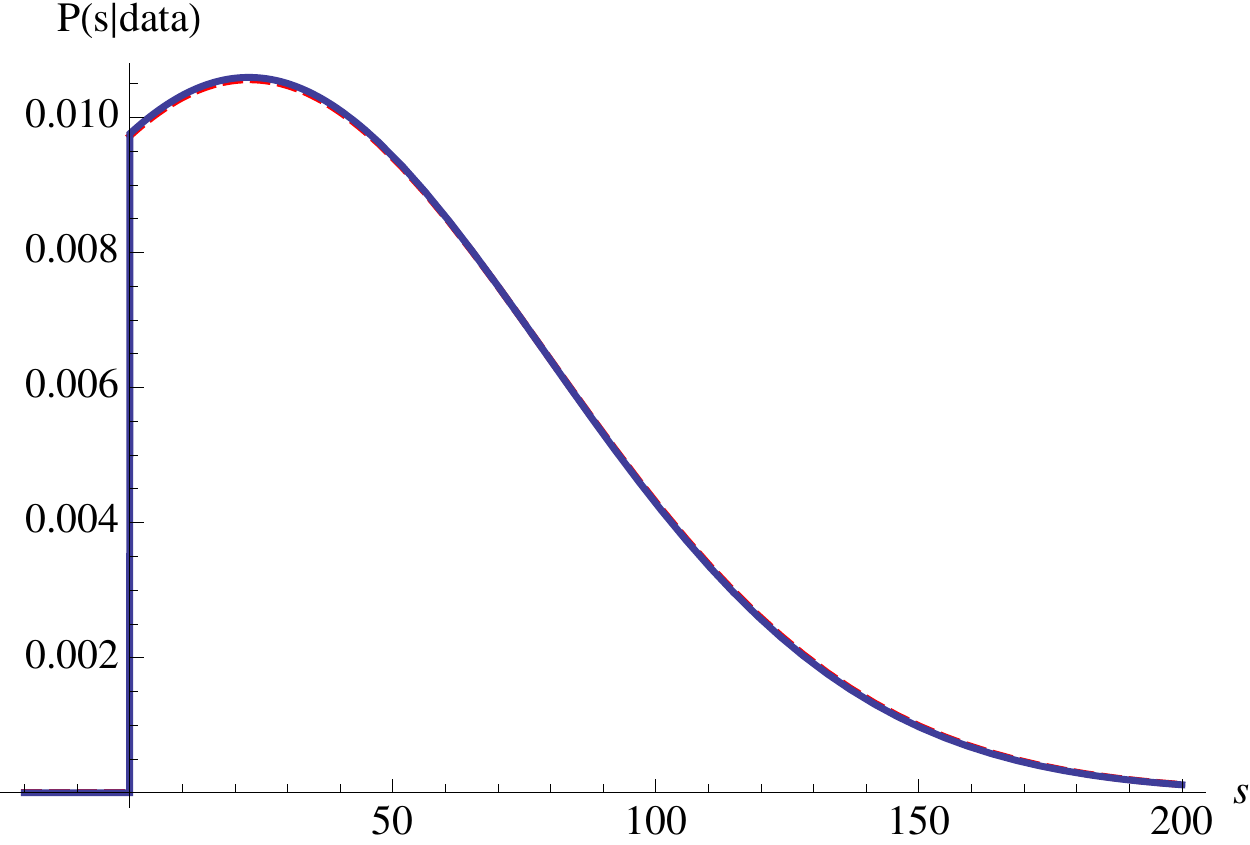} \label{fig:convPosteriorRMSS}}
\caption{Dashed red: The posterior $p(s|\text{data})$ without any systematic uncertainty.  Solid blue: The same posterior after convoluting systematic uncertainty.   Left: Using the systematic uncertainty of the example in Section~\ref{sec:convMean}.  Right: Using the systematic uncertainty of Section~\ref{sec:convRMS}.}
\end{center}
\end{figure}

\subsection{Example of uncertainty in signal width}
\label{sec:convRMS}

In this example we follow the same steps as in Section~\ref{sec:convMean}, except that we formulate {\tt f[n]} at the beginning in a different way.  Here we wish $n$ to parametrize some uncertainty in the width of signal which is known to be Gaussian with mean equal to 15 and width somewhere near 3.  Here is the only different command:
\begin{lstlisting}
f[n_] := A*Table[Exp[-(15 - i)^2/(2*(3*n)^2)], {i, 1, nBins}]/
    Sum[Exp[-(15 - i)^2/(2*(3*n)^2)], {i, 1, nBins}]];
\end{lstlisting}
Figure~\ref{fig:convSig2} shows some examples of this {\tt f[n]}.

The posterior is assumed the same as in the previous example.  The resulting $p(s,n|\text{data})$ is shown in Fig.~\ref{fig:convPosteriorRMSSN}, and the final $p(s|\text{data})$ in Fig.~\ref{fig:convPosteriorRMSS}.

\begin{figure}
\begin{center}
\includegraphics[width=0.45\textwidth]{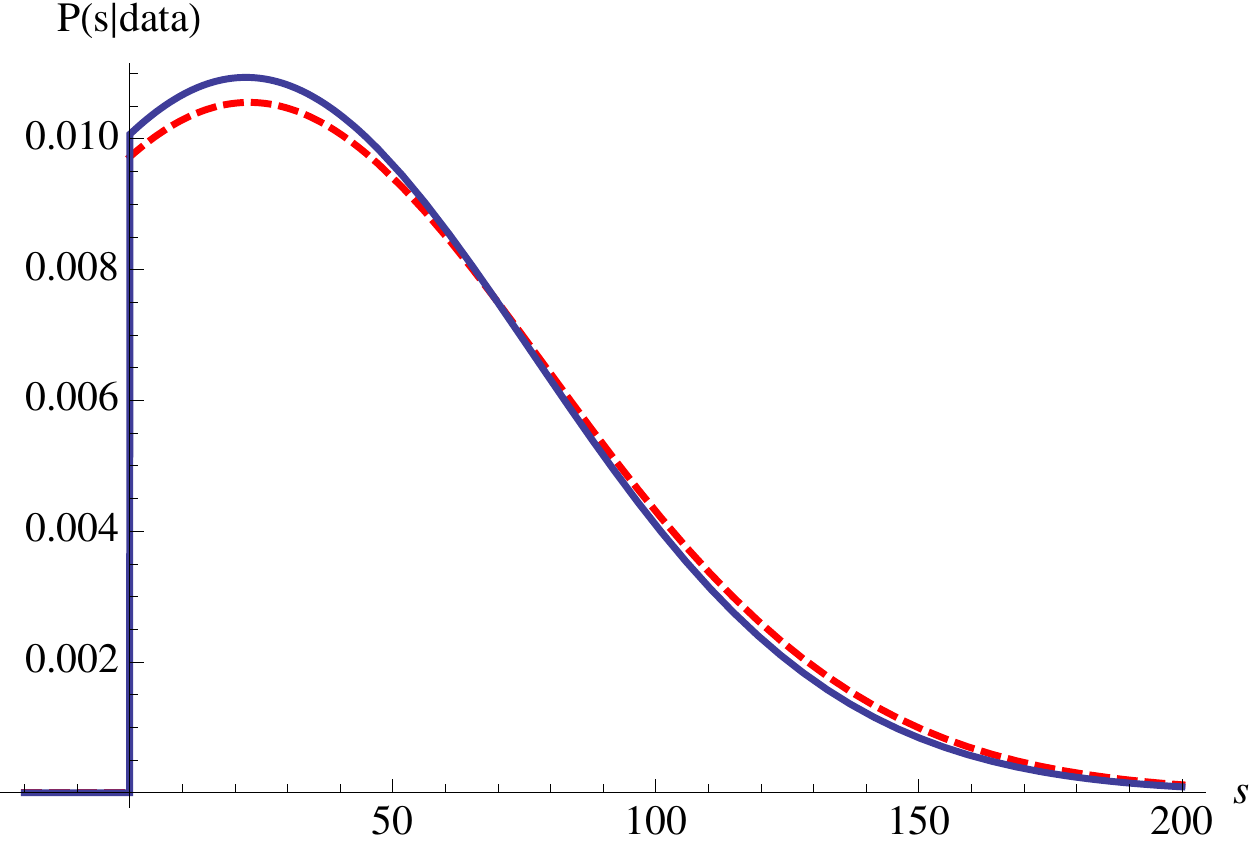}
\caption{Same as Fig.~\ref{fig:convPosteriorRMSS}, except that this time the systematic uncertainty of Section~\ref{sec:convRMS} is convoluted using a prior which does not constrain $n$ to be Gaussian-distributed around 1$\pm$0.1, but gives $n$ equal probability to be anywhere between 0.5 and 1.5. \label{fig:convFlatRMS}}
\end{center}
\end{figure}

Interestingly, the effect of this uncertainty is much smaller than the uncertainty of Section~\ref{sec:convMean}.  Not only it is much smaller, but it goes in the opposite direction:  it makes the upper limit slightly stricter than if we had no uncertainty at all.  Specifically, the 95\% upper limit moves from 131.315 to 130.825.  This is admittedly a minuscule improvement, but it is possible to find an example where the improvement is noticeable.  For example, if instead of the prior of Section~\ref{sec:convMean} we use a ``box'' prior in $n$:
\begin{lstlisting}
Prior[s_, n_] := UnitStep[s] * UnitStep[n-0.5]*UnitStep[1.5-n]
\end{lstlisting}
then the effect of this width systematic uncertainty is more visible (Fig.~\ref{fig:convFlatRMS}), and it changes the 95\% upper limit to 126.7, which is a more clear improvement.

Many people are under the impression that systematic uncertainties have to always make limits worse, because ``less information has to make things worse'', or something along these lines.  This is a verbal over-simplification of the actual mathematical procedure.  Systematic uncertainty is not only ``less information''; it is also ``more possibilities''.  Upper limits get worse (looser) when the data show an excess, and better (stricter) when there is a deficit.  If we have an excess, and no uncertainty whatsoever, we are in a situation that disfavors the limit, and there is no chance the situation is any different.  But if some systematic uncertainty is introduced, it might allow some scenarios where the situation is more favorable.  If we average out all scenarios, which is what the convolution of eq.~\ref{eq:posteriorNuis1} does, then the limit might improve.  Empirically, this doesn't happen very often, but it has been observed several times, in numerous analyses, including the numerical example above.

\section{Computing the coverage of limits}
\label{sec:coverage}

This section will please readers who like Frequentist limits.  The ``holy grail'' in Frequentist limits is {\em coverage}.  Frequentist constructions provide (or should provide at least) intervals of specific coverage, a typical choice in High Energy Physics being 95\%.  Intervals of coverage 95\% are called ``95\% confidence intervals'' (CIs).  

\paragraph{What is coverage?}  To understand that, one needs first to realize that, even if the laws of Nature don't change, a different observation of Nature would result in different data;  there are random fluctuations.  Both Bayesian inference and Frequentist constructions use data as input, so, their outputs (PDFs and CIs) are also subject to statistical fluctuation.  If the POI has some value ($v$), and we collect many (in principle infinite) independent datasets, and we compute an interval using each one of these datasets, we will find $v$ within our interval with frequency $c$.  This number ($c$) is the coverage of the interval.  It is a statistical property of the interval, and the procedure used to determine it.  Obviously, the coverage may depend on $v$, and on the procedure used to find the interval.

Coverage is not only a property of Frequentist intervals;  Any interval, however it is defined, has some coverage.

To compute the coverage of a Bayesian credibility interval, we can write a loop which repeatedly creates pseudo-data that are consistent with some assumed value of the POI, repeats the Bayesian limit-setting procedure, and in the end count how many times the assumed value of the POI was within the interval.  

The following lines compute the coverage of an upper limit with 95\% credibility
\begin{lstlisting}
b = {21000., 14000., 10000, 7100., 4800., 3400., 2300., 1600., 1100., 740., 500, 350., 230., 160., 100, 70., 46., 30., 20., 13., 8.2, 5.2, 3.2, 2.0, 1.2, 0.71, 0.42, 0.24, 0.13, 0.074};
f = {0, 0.0000105, 0.000335, 0.000485, 0.00015, 0.0008, 0.00115, 0.00425, 0.0022, 0.0034, 0.00495, 0.0055, 0.0095, 0.018, 0.0185, 0.028, 0.085, 0.21, 0.085, 0.0125, 0.0044, 0, 0.0000105, 0, 0, 0.000335, 0, 0, 0, 0};
nBins = Length[b]
A = Total[f]
L[s_] := Exp[Sum[d[[i]]*Log[Max[0, b[[i]] + s*f[[i]]]], {i, 1, nBins}] - s*A]/L0
Prior[s_] := UnitStep[s]
Posterior[s_] := L[s]*Prior[s]/NormConst;
v = 100;
nPseudo = 1000;
cred = 0.95;
answers = Table[0, {i, 1, nPseudo}];
Do[
 d = Table[Random[PoissonDistribution[b[[i]] + v*f[[i]]]], {i, 1, nBins}];
 L0 = Exp[Sum[d[[i]]*Log[b[[i]]], {i, 1, nBins}]];
 NormConst = NIntegrate[L[s]*Prior[s], {s,-Infinity,Infinity}];
 integ = NIntegrate[Posterior[s], {s, -Infinity, v}];
 If[integ < cred, answers[[i]] = 1], 
 {i, 1, nPseudo}]
N[Total[answers]/nPseudo]
\end{lstlisting}

\begin{description}
\item[Line 1:] Define the background in each bin.  Same as in Section~\ref{sec:poisson}.
\item[Line 2:] Define the signal distribution.  Same as in Section~\ref{sec:poisson}.
\item[Line 3:] Number of bins.  {\tt nBins} is 30 in this example.
\item[Line 4:] The acceptance to the signal, given by eq.~\ref{eq:acceptance}.  In this example $A \simeq 0.49$.
\item[Line 5:] Define the likelihood function.
\item[Line 6:] Define a flat prior for $s>0$.  It can obviously change, and the coverage will have some dependence on the prior, since the prior is part of the procedure that defines the Bayesian credibility interval.
\item[Line 7:] Define the formula for the posterior.
\item[Line 8:] The assumed amount of produced signal.  Variable {\tt v} corresponds to the $v$ used above, in the definition of coverage.  This is the amount of signal that will be added to the background to generate each set of pseudo-data, in line 13.
\item[Line 9:] Define how many iterations to make in the loop which starts in line 12 and ends in line 18.  A large number of iterations will lead to a more precise estimation of the actual coverage.
\item[Line 10:] Define the credibility level of the upper limit whose coverage will be estimated.  0.95 means 95\% credibility level.
\item[Line 11:] Initialize an array of {\tt nPseudo} answers.  The elements initially are all 0, and some of them will turn into 1 inside the loop, in line 17.  Each element represents the output of a set of pseudo-data.  If the element is 0, it means that the interval failed to contain the true POI value ($v$).  If it is 1, it means that the interval succeeded to contain $v$, namely the upper limit is a number greater than $v$.
\item[Line 12:] Starting the loop.
\item[Line 13:] Define the data, which consist of Poisson fluctuations of the content of each bin, with mean equal to the background of the bin, plus the signal events that would end up in the bin if $v$ signal events were produced.  Clearly, {\tt d} are data consistent with the hypothesis that $v$ signal events are produced.
\item[Line 14:] Calculate the constant {\tt L0} which is introduced in line 5 to make {\tt L[s]}  easier the handle numerically.
\item[Line 15:] Compute the normalization constant which normalizes the posterior defined in line 7.
\item[Line 16:] Compute $\int_{-\infty}^v p(s|\text{data})\;ds$, and store it in variable {\tt integ}.
\item[Line 17:] If {\tt integ} is less than {\tt cred}, then register the value 1 in the {\tt answers} array, in the position that corresponds to the current pseudo-data set.  The logic is that, if {\tt integ} is less than {\tt cred}, then the upper limit with credibility {\tt cred} can't be but some number greater than $v$.  That's obvious, since the upper limit is defined as the $x$ which satisfies $\int_{\-infty}^x p(s|\text{data})\;ds = {\tt cred}$.  This trick allows us to know if the interval covers $v$, without really computing the interval, which would be a more CPU-expensive computation.
\item[Line 18:] The loop closes, after {\tt nPseudo} iterations.
\item[Line 19:] Out of the {\tt nPseudo} trials, some have succeeded, in the sense that the interval covered the actual POI value ($v$).  We can count these successes by summing the elements of the {\tt answers} array.  Dividing by {\tt nPseudo}, we get an estimator of the success rate, which is, by definition, the coverage.
\end{description}

Running the above code, with the numbers given, returned coverage 0.960.
Smaller values of $v$ result in larger coverage, and when $v$ increases the coverage asymptotically becomes equal to credibility, namely 0.95.  This is true for any prior one may assume, and there are some special non-informative priors which make the convergence faster.

\section{Summary}

It has been shown how to compute posterior PDFs and limits to any arbitrary signal in the most common case of Poisson-distributed data (Section~\ref{sec:poisson}) and in the case of binomially distributed data (Section~\ref{sec:binomial}).

The treatment is described for signals that are not simply additive to the background, but interfere with it (Section~\ref{sec:nonAdd}).

It was then shown how to combine datasets in the most general case where the datasets are coming from dissimilar experiments and dissimilar observables (Section~\ref{sec:combo}).  

Then the case of simultaneous estimation of multiple POIs was shown in Section~\ref{sec:multidim}.

All the above computations assumed no systematic uncertainties, until Section~\ref{sec:systematics}, where the principle was laid out to perform convolution of systematic uncertainties, and two complete examples where shown.

Finally, Section~\ref{sec:coverage} shows the way to compute the coverage of a Bayesian upper limit, which can be interesting to someone who, being used to Frequentist limits, may appreciate coverage.

Emphasis has been given to the practical implementation of all computations, and remarks have been made to gain some insight in the results.

\section{Acknowledgements}

G.C.\ thanks the theorists Michele Redi and Mads Frandsen for their encouragement to proceed with this work, expecting it to be welcomed with interest by many theorists and experimentalists.

He also thanks his ATLAS collaborators, Glen Cowan, Alex Read, Eilam Gross, David Adams, and Diego Casadei, for all our discussions\footnote{Thanking people for discussing does not imply that they necessarily endorse everything written here.  Responsibility lies with the author.}.

If you spot errors, thank you in advance for informing the author (\url{gchouda@alum.mit.edu}).  
 
Finally, if you learned something useful that you can apply in your phenomenological or experimental research, please cite this document, to acknowledge the author for his effort.  Thank you.

%%%%%%%%%%%%%%%%%%%%%%%%%%

\end{document}